\documentclass[aip, preprint, cha]{revtex4-1}
\usepackage{00_My_Packages}

\begin{document}

\newcommand{\QgsGroundTemp}[1][]{%
    \ifthenelse{\isempty{#1}}{\delta T_{\mathrm{g}}}{\delta T_{\mathrm{g},\,#1}}
}

\newcommand{\QgsAtmTemp}[1][]{%
    \ifthenelse{\isempty{#1}}{\delta T_{\mathrm{a}}}{\delta T_{\mathrm{a},\,#1}}
}

\newcommand{\QgsBaroclinic}[1][]{%
    \ifthenelse{\isempty{#1}}{\theta_{\mathrm{a}}}{\theta_{\mathrm{a},\,#1}}
}

\newcommand{\QgsBarotropic}[1][]{%
    \ifthenelse{\isempty{#1}}{\psi_{\mathrm{a}}}{\psi_{\mathrm{a},\,#1}}
}

\newcommand{\QgsFriction}{k_d}
\newcommand{\QgsInsolation}{C_{\mathrm{g}, 1}}
\newcommand{\QgsAtmInsolation}{C_{\mathrm{a}, 1}}
\newcommand{\corriolis}{f_0}
\newcommand{\emissivity}{\varepsilon_{\mathrm{a}}}

\newcommand{\lNorm}{\mathrm{L}^2}

\newcommand{\Trajectory}{\mathbf{x}}

\newcommand{\UPOperiod}{T}
\newcommand{\system}{S}
\newcommand{\IC}[1][0]{\mathbf{x}_{#1}}
\newcommand{\systemFlow}[1][t]{\system^{#1}}
\newcommand{\UPO}[2]{\overline{\system^{#1}_{#2}}}
\newcommand{\Monodromy}{\bm{M}}
\newcommand{\CombinedVector}[1][i]{w_{#1}}

\newcommand{\stableSS}{S}
\newcommand{\stableSSu}{s}
\newcommand{\unstabSS}{U}
\newcommand{\unstabSSu}{u}

\newcommand{\shadowing}{\Omega}
\newcommand{\Landing}[1][L]{U_{\mathrm{L}}}
\newcommand{\PreTransitionSet}[1][L]{U_{\mathrm{L}}}
\newcommand{\TransitionSet}[1][L]{T_{\mathrm{L}}}

\newcommand{\UPOIndexFunc}{\text{ind}}

\newcommand{\transpose}[1]{#1^{\mathrm{T}}}
\newcommand{\idMat}{\mathrm{I}}
\newcommand{\jac}[1][t]{J^{#1}}

\newcommand{\Deriv}[2]{\frac{\mathrm{d} #1}{\mathrm{d} #2}}
\newcommand{\partialDeriv}[2]{\frac{\partial #1}{\partial #2}}
    \title{Using Unstable Periodic Orbits to Understand Blocking Behaviour in a Low Order Land-Atmosphere Model}

\author{Ois{\'i}n Hamilton}
\email{oisin.hamilton@meteo.be}
\affiliation{Royal Meteological Institute of Belgium}
\affiliation{UCLouvain --- Earth and Life Institute}

\author{Jonathan Demaeyer}
\affiliation{Royal Meteological Institute of Belgium}

\author{Michel Crucifix}
\affiliation{UCLouvain --- Earth and Life Institute}

\author{Stéphane Vannitsem}
\affiliation{Royal Meteological Institute of Belgium}

\date{\today}

\begin{abstract}
	Unstable Periodic Orbits (UPOs) were used to identify regimes, and transitions between regimes, in a reduced-order coupled atmosphere-land spectral model. In this paper we describe how the chaotic attractor of this model was clustered using the numerically derived set of UPOs. Using continuation software, the origin of these clusters were also investigated. The flow of model trajectories can be approximated using UPOs, a concept known as shadowing. Here we extend that idea to look at the number of times a UPO shadows a model trajectory over a fixed time period, which we call cumulative shadowing. This concept was used to identify sets of UPOs that describe different life cycles of each cluster. The different regions of the attractor that were identified in the current work, and the transitions between these regions, are linked to specific atmospheric features known as atmospheric blocks.
\end{abstract}

\maketitle

\begin{quotation}
	Unstable Periodic Orbits (UPOs) are distributed densely in a chaotic attractor, meaning that any trajectory is finitely close to a UPO at any given time. They therefore provide a means of describing properties of the attractor. In this paper we use UPOs to classify the regimes of a simple climate model and to identify the transitions between regimes. The origin of the regimes is also investigated using continuation software to track key UPOs through parameter space. The regimes of this model correspond to atmospheric blocking behaviour, which is linked with heat waves and cold snaps in the real atmosphere. Currently the physical mechanisms of atmospheric blocking are still not fully understood and simple climate models can play a useful role in identifying key physical behaviour.     
\end{quotation}
    \section{Introduction}\label{sec:intro}
    Climate change will increase the intensity, frequency, and duration of heat waves~\citep{intergovernmentalpanelonclimatechangeipcc2023}. The atmosphere, while characterised by chaotic behaviour, can exhibit some level of repetition over time spans of weeks or months. This quasi repeating behaviour is called Low Frequency Variability (LVF). One example of LFV is midlatitude atmospheric blocking (from here referred to as blocks or blocking). Blocking causes the jet stream to be deflected, and in the midlatitudes this deflects the eastward moving low pressure systems, while the area in the trough of the deflection experiences persistent weather~\citep{boyd2015}. This usually leads to heat waves and cold snaps in the midlatitudes. Numerical weather prediction methods still struggle with predicting the onset and decay of such events~\citep{davini2020}. This is believed to be caused by missing subgrid processes and regional specific features. However there is still no clear consensus on the exact key physical processes that cause blocking~\citep{badza2024, davini2020, pinheiro2019}. Therefore, to fully understand the nature of blocking, it should also be studied from a conceptual point of view.

    One way to tackle this problem conceptually is to use reduced order models to look at new approaches to describe and predict the dynamical properties of the atmosphere. It is at the core of the concept of ``bottom-up" analysis, which tries to identify fundamental dynamical mechanisms from fluid mechanics, thermodynamics, and dynamical system theory. 
    Low-order models build upon a minimal set of key physical ingredients believed to be at the origin of the mechanisms of interest and can therefore act as a testing ground to investigate new concepts for understanding and predicting LFV (see~\citet{vautard1988, mak1989, nauw2001, vannitsem2015, vannitsem2021} for some examples). 
    These models provide a cheap and flexible platform upon which to test concepts, and can be simple enough to understand the dynamics which produce the climatological behaviour. The hope is that the mechanisms identified through the low order models survive in more complex models and can help to understand and characterise this behaviour.

    Consistent with this strategy we aim study the onset and decay of blockings from a dynamical systems perspective, by using the \textsf{qgs} (quasi-geostrophic spectral) model framework~\citep{demaeyer2020} which we describe below. Our objective is to link features of the state space of the attractor, to the climatology of the model, with the aim of then describing the atmospheric behaviour that we see from the dynamics of the model. 
    
    Weather regimes and LFV have been studied from a dynamical systems perspective in reduced order models since the early days of numerical weather prediction~\citep{lorenz1963, veronis1963}. However, caution needs to be taken when interpreting results from low order models. On top of the obvious fact that these models only reproduce select behaviour of the real atmosphere, they can also be highly susceptible to which wave modes (or model resolution) are included. This means that the resulting chaotic attractor, and in turn repeating behaviour in the attractor, can differ wildly~\citep{cehelsky1987}. Convergence of the structure of the attractor with model resolution should therefore be checked~\citep{decruz2016}.

    Work on reduced order models led to the hypothesis that persistent weather phenomena, and different weather regimes could be explained by proximity to the trajectory of different fixed points with differing stability properties~\citep{charney1979, charney1980, reinhold1982}. The idea has since been expanded to look at areas of periodically repeating behaviour~\citep{kazantsev1998}, and other regions of quasi-stationary, or very slowly altering behaviour~\citep{badza2024}. Our aim is to better understand how trajectories evolve through these systems, and how parameter changes will alter the chaotic attractor. These questions can be addressed using periodic orbit theory, which considers the periodic orbits of a dynamical system as building blocks of its attractor, helping to construct a skeleton of the dynamics~\citep{cvitanovic2020}.
    
    It was conjectured by Poincar\'e in the 19th century that chaotic systems could be approximated by periodic motions that are densely distributed within the attractor~\citep{birkhoff1927}. Unstable Periodic Orbits (UPOs) are closed loops in state space that provide a kind of topological invariant characterisation of the attractor~\citep{cvitanovic1988}. This means that they preserve topological relations between periodic orbits, such as their relative inter-windings, allowing for at least a partial organisation of the state space. The use of these orbits to investigate the state space has since been called periodic orbit theory~\citep{cvitanovic2020}, and has been used in a wide range of fields to help describe properties of the attractor~\citep{grebogi1988, pughe-sanford2023, crane2025}. 

    Though numerically finding all of the UPOs in the chaotic attractor of a system is impossible as they form a dense set, it has been found that usually a limited number of them - likely of low period \citep{hunt1996, hunt1996a} - can be used to describe the predominant behaviour of the system. In simple systems~\citet{maiocchi2022} found that UPOs could help predict transitions between wings of the Lorenz 1963 attractor, among other important stability properties of the attractor. This work was expanded to the Lorenz 96 model in \citet{maiocchi2024}, where the authors suggest that Lyapunov analysis should be complimented with the use of UPOs to provide a global picture of the attractor.
    
    UPOs have also been used in reduced order models of the climate. Using a barotropic ocean model \citet{kazantsev1998} found that even with only a few low order UPOs, some key properties of the attractor, such as the PDF of model trajectories, could be predicted. This work was built on by \citet{selten2004} using the Lorenz 84 model, and later by \citet{gritsun2013} using a barotropic atmosphere model. 
    
    \citet{kazantsev2001} investigated the impact of forcing on the model, by analysing the impact of the forcing on a small set of UPOs ($\sim20-30$). This method allowed them to find the magnitude of forcing that caused the largest change in the averaged dynamics, allowing them to identify which forcings would have the largest impact on the climate of the model. \citet{gritsun2008}, using a barotropic atmosphere model, attempted to use UPOs to calculate the response of the system to small external forcings. They concluded that the method could be utilised for understanding local sensitivity on specific orbits, but likely cannot be used to understand the global behaviour. This is a result of the computational cost in finding the UPOs, and the fact that it is impossible to know in more complex systems if all UPOs up to a period of $\UPOperiod$ have been found.
    
    \citet{lucarini2019a} extended the approach of using UPOs to structure the state space in an intermediate complexity model. They found that blockings occur when the model trajectory is in the neighbourhood of a particular set of UPOs, and that these UPOs generally have higher instability compared to the set of UPOs that describe zonal flows. This result could explain the local instability associated with blocking events. However, these promising results from periodic orbit theory become challenging to use in high Lyapunov dimension systems (the number of ordered Lyapunov exponents that can be summed and the total to be greater than 0)~\citep{crane2025}. Therefore the requirement for the orbits to be closed has been relaxed in some studies and extended to quasi recurrent patterns in the attractor~\citep{badza2024}.

    In this paper we use UPOs to provide an explanation to the structure of the chaotic attractor of a reduced-order land-atmosphere model, and explain the persistence seen in different parts of the attractor. We use the UPOs to also define transition regions in the attractor. The onset and decay of blockings is also investigated to see whether collections of UPOs can be used to identify these events. In Section~\ref{sec:methods_sub:model} we introduce the model used, and then discuss, in Section \ref{sec:methods_sub:finding_upos}, how we numerically found the UPOs. In Section~\ref{sec:results_sub:key_upos} we describe the UPOs found in the model and how they appear to centre around two clusters, then in Section~\ref{sec:results_sub:cluster_origin} we describe the origin of these clusters for certain parameter values. In Section~\ref{sec:results_sub:clustering} we describe how we use the UPOs to cluster the attractor, the transitions between these clusters (Section~\ref{sec:results_sub:transitions}), and lastly how we can use the UPOs to predict transitions (Section~\ref{sec:results_sub:upo_shaddowing}).

    \section{Methods}\label{sec:methods}
\subsection{Model}\label{sec:methods_sub:model}
    The \textsf{qgs} (quasi-geostrophic spectral) model framework~\citep{demaeyer2020} consists of a collection of reduced-order models where the atmosphere is composed of two layers coupled to a choice of different bottom layers with different degrees of interactions. The framework also allows for a flexible spectral resolution. In this paper we introduce a significantly modified version of the \textsf{qgs} framework, which uses a symbolic Python library (SymPy) to return the model equations including parameter values~\citep{meurer2017}. This is instead of fixing the numerical values of parameters, as was the case in the original framework, when the equations of the model are generated.

    In the model, the atmosphere component consists of a channel quasi-geopstrophic atmosphere on a $\beta$-plane. The model aims to simulate large scale dynamics of the midlatitudes, and approximately covers the latitudes $\pm20^\circ$ of the central latitude at $50^\circ$. 
    
    In this study we are coupling this atmosphere to a land bottom layer, and the two components are coupled through wind stress and a thermal energy balance scheme~\citep{li2018}. The equations describing the evolution of the barotropic ($\QgsBarotropic[]$) and baroclinic ($\QgsBaroclinic[]$) streamfunctions are:
    
    \begin{equation}
        \begin{aligned}
            \partialDeriv{}{t}\left(\nabla^2 \QgsBarotropic{}\right)+J\left(\QgsBarotropic{},~\nabla^2\QgsBarotropic{}\right) + J\left(\QgsBaroclinic{},~\nabla^2\QgsBaroclinic{}\right) &+ \frac{1}{2}J\left(\QgsBarotropic{} - \QgsBaroclinic{},~\corriolis h/H_{\mathrm{a}}\right)+\beta\partialDeriv{\QgsBarotropic{}}{x}\\
            &= -\frac{\QgsFriction}{2}\nabla^2\left(\QgsBarotropic{}-\QgsBaroclinic{}\right)\\
            \partialDeriv{}{t}\left(\nabla^2 \QgsBaroclinic{}\right)+J\left(\QgsBarotropic{},~\nabla^2\QgsBaroclinic{}\right) + J\left(\QgsBaroclinic{},~\nabla^2\QgsBarotropic{}\right) &- \frac{1}{2}J\left(\QgsBarotropic{} - \QgsBaroclinic{},~\corriolis h/H_{\mathrm{a}}\right)+\beta\partialDeriv{\QgsBaroclinic{}}{x}\\
            &=-2\QgsFriction^{'}\nabla^2\QgsBaroclinic{} + \frac{\QgsFriction}{2}\nabla^2\left(\QgsBarotropic{}-\QgsBaroclinic{}\right) + \frac{\corriolis}{\Delta p}\omega
        \end{aligned}
    \end{equation}
    where $\QgsFriction$ is the friction between the ground and the atmosphere, $\QgsFriction^{'}$ is the internal friction of the atmosphere, $h$ controls the height of the orography, and $H_{\mathrm{a}}$ is the characteristic depth of the atmospheric layers. 

    The energy balance between the ground and the atmosphere heat is modelled as:

    \begin{equation}
        \begin{aligned}
            \gamma_{\mathrm{a}}\left(\partialDeriv{T_{\mathrm{a}}}{t} + J\left(\QgsBarotropic{},~ T_{\mathrm{a}}\right)-\sigma \omega \frac{p}{R}\right) &= -\lambda\left(T_{\mathrm{a}}-T_{\mathrm{g}}\right) + \emissivity\sigma_{\mathrm{B}}T^4_{\mathrm{g}} - 2\emissivity\sigma_{\mathrm{B}}T^4_{\mathrm{a}} + R_{\mathrm{a}}\\
            \gamma_{\mathrm{g}}\partialDeriv{T_{\mathrm{g}}}{t} &= -\lambda\left(T_{\mathrm{g}}-T_{\mathrm{a}}\right) - \sigma_{\mathrm{B}}T^4_{\mathrm{g}} + \emissivity\sigma_{\mathrm{B}}T^4_{\mathrm{a}} + R_{\mathrm{g}}\\
        \end{aligned}
    \end{equation}
    here $\gamma_{\mathrm{a}}$ and $\gamma_{\mathrm{g}}$ are the heat capacities of the atmosphere and the land, $\sigma$ is the static stability of the atmosphere, $\lambda$ combines both latent and sensible heat fluxes and $\sigma_{\mathrm{B}}$ is the Stefan-Boltzmann constant. In the model the longwave radiation terms are linearised. Lastly, to reduce the number of model variables, a relation between the baroclinic streamfunctions and the atmospheric temperature is used ($\QgsAtmTemp{}=2\corriolis\QgsBaroclinic{} / R$, where $R$ is the ideal gas constant). 
    
    In this paper we configure the model as described in~\citet{xavier2024}, where we set the model resolution to be of wavenumber 2 in the $x$ and $y$ direction, resulting in 10 basis functions. Here the land and atmosphere use the same basis. The key parameter values used are given in Table~\ref{tab:key_parameters}.

    The new \textsf{qgs} framework (qgs v1.0) provides two key improvements over the previous version. The first is that selected parameters can now be left as an undefined variable~\citep{meurer2017} when generating the model tendencies, i.e. the model tendencies become a function of the selected parameters (in addition to being a function of the state space). Second, the model equations can now be returned in any desired programming language. Combined together, these two features allowed us to create a pipeline that directly feeds the model equations into continuation software to study the dynamics of the model.\footnote{The release of the v1.0 of the \textsf{qgs} framework is in preparation and should be available online before the end of the review process.}

    \begin{table}
        \begin{tabular}{|c|l|l|}
            \hline
            Parameter & Value & Description\\
            \hline
            $\emissivity$ & 0.76 & Atmospheric emissivity\\
            $\QgsFriction$ & 0.085 [non dim] & Friction between the atmosphere and the ground\\
            $\QgsFriction^{'}$ & 0.02 [non dim] & Internal friction between the two layers of the atmosphere\\
            $\QgsInsolation$ & 300\si{\watt\per\square\meter}& Solar forcing\\
            $\QgsAtmInsolation$ & $\frac{4}{10}\QgsInsolation\si{\watt\per\square\meter}$ & Solar forcing\\
            \hline
        \end{tabular}
        \caption{Unless otherwise stated, the model is set with the above parameter values.}
        \label{tab:key_parameters}
    \end{table}

\subsection{Numerically Finding UPOs}\label{sec:methods_sub:finding_upos}
    In this paper we have used two methods for finding UPOs: the Newton-Raphson method and the continuation software AUTO~\citep{doedel2012}. The Newton-Raphson method~\citep{mestel1987, parker1989} aims to solve the equation $\systemFlow(\IC)=\IC$, where the unknowns are the starting location in state space ($\IC$), and the time $t$. 
    This method has been described in numerous studies to numerically calculate UPOs~\citep{farantos1995, saiki2007, abad2011, barrio2015, maiocchi2022}. Following~\citet{gritsun2008} we have used the second order tensor method, which aids in finding UPOs in this system where there are a large number of close to zero Lyapunov exponents~\citep{gritsun2010}. In this case, due to the long periods of UPOs, the region of convergence can be very small and the second order method, though approximately 10 times more computationally expensive, can aid with finding UPOs~\citep{crofts2009} in these circumstances. The algorithm used is described in Section~\ref{app:numerical_methods}. 
    
    We also used the software AUTO to find UPOs via continuation across parameters. This was done by first numerically finding the fixed points of the model. We tracked these fixed points for several key parameter values. At any Hopf bifurcation points we then branched and followed the periodic orbit branches. We continued to follow branches at branching points or period doubling from these periodic branches. This process was done for up to 50 branches deep. This process was aided by the auto-AUTO Python package~\citep{demaeyer2025, demaeyer2025_zenodo}. This package acts as a layer on top of AUTO, and automates the branching and identifies which branches to follow.

\subsection{Initial Conditions}\label{sec:methods_sub:ic}

    The Newton-Raphson method can be very sensitive to initial conditions. This means that which UPO you end up on, or whether you end up on a UPO at all, is highly dependent on the initial conditions fed into the algorithm. The initial conditions refer to the starting location of the closed loop in state space, and the initial guess at the period of the closed loop. If the state space is of dimension $n$ the initial condition is of dimension $n+1$. In this study we used the following four methods to guess initial conditions:

    \subsubsection*{Random Initial Conditions} 
    The initial conditions, the starting location in space and the period, were chosen randomly. We used a uniform distribution, where the bounds of the uniform distribution were set by finding the bounds of the attractor. Here we set the maximum period to be 225 days. This upper bound on the period was set as the probability of finding longer period UPOs is low.  

    \subsubsection*{Near Misses} 
    We first ran a very long trajectory. We then passed a sliding window of length 225 days over this trajectory, and by calculating a norm between every point within this sliding window, we determined points ($\IC$) where after some time $t_i$ the trajectory passes sufficiently close to itself again: $\left\|\systemFlow[t_i](\IC) - \IC\right\|<\varepsilon$. This provided us with a guess of the period ($\UPOperiod_0=t_i$) and the starting point in space $\IC$.

    \subsubsection*{Perturbed UPOs} 
    Starting from a UPO we take a random point on the UPO and perturb the point and the period. We use this perturbed point as a new initial condition and feed this to the algorithm. This utilises the fact that UPOs should be dense in the attractor, so there should be another UPO living close to the given one. Usually this will result in the method producing the original UPO, so a filter needs to be used to ignore the result if it does not produce a new UPO. 

    \subsubsection*{UPO Continuation} 
    UPOs could be found for certain parameter values more easily than others, in certain regions of the state space. This can occur when the largest Floquet multiplier of the UPO decreases with a change in parameter, or when the period of the UPO decreases with a change in the parameter. 
   
    This method takes advantage of this, by first finding a UPO at parameter values where it is easier to do so. Continuation software is then used to track the UPO back to the sought after parameter values. In our case, we used the work of~\citet{xavier2024} to understand for which parameter values the largest Lyapunov exponent was close to zero, suggesting that the flow is diverging more slowly at these values. We then used the Newton-Raphson method to find UPOs for these parameter values and tracked these branches back to the original values using AUTO. This method will fail when a the branch does not extend to the desired parameters either due to a fold or the continuation software not being able to track the branch.

    \section{Results}\label{sec:results}
\begin{figure}
        \centering
        \includegraphics[width=\linewidth]{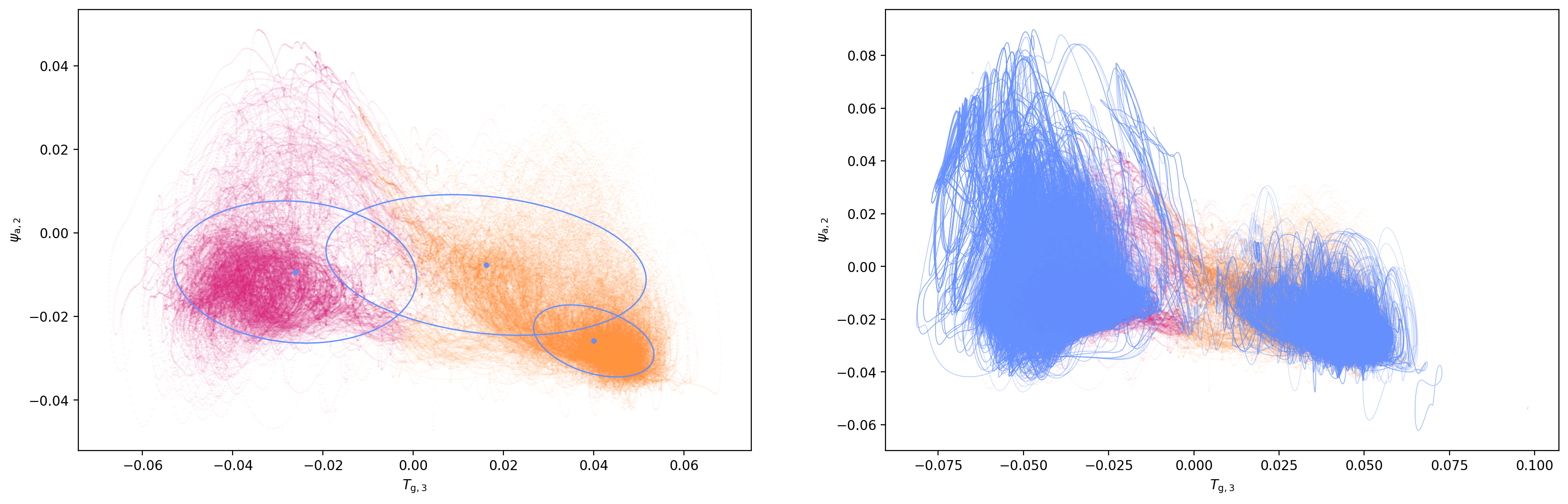}
        \caption{The chaotic attractor of the model, where the two colours (pink and orange) show the clustering of the attractor using the distance from UPOs. On the left, the Gaussian mixture clustering clusters are shown in blue, calculated as described in~\citet{xavier2024}. On the right we show the UPOs projected on top of the attractor.}
        \label{fig:cluster}
    \end{figure}

    When the chaotic attractor is projected onto the variables $(\QgsGroundTemp[3], \QgsBarotropic[2])$, shown in Figure~\ref{fig:cluster}, two main clusters are visually present. The clusters seen on the left and right designate regions where averaged behaviour displays, on average, atmospheric blocking behaviour, driven by the orography and the temperature in the ground.
    
    Projecting the 30 dimensional state space on to these variables makes intuitive sense for studying atmospheric blocking regimes. In this low order model we expect blocks to look like a classic $\Omega$-block, where we have one large high pressure with a low pressure on either side. In addition we expect the ground temperature anomaly to create a higher geopotential anomaly. These variables present the strength of the barotropic atmosphere and ground temperature variables on the modes $F_2=2\cos(nx)\sin(y)$ and $F_3=2\sin(nx)\sin(y)$. Large magnitudes of these variables are likely to signify regimes that display, on average, blocking in our model. The transitions between these clusters display zonal atmospheric behaviour or transitions between leeward and windward blocking regimes of the orography.

\subsection{Key Unstable Periodic Orbits}\label{sec:results_sub:key_upos}
    We found 6818 UPOs in total, using the methods described in Section~\ref{sec:methods_sub:finding_upos}. These UPOs are primarily centred on top of the two dense clusters, designating the two key weather regimes described above. The Newton method primarily found UPOs in regions where model trajectories remain for extended time periods. This is likely due to adequate initial conditions being easier to stumble across. We could not find any UPOs that link the two central clusters. This is likely because such periods would either have periods longer than the maximum searched for (set to 225 days as beyond this the algorithm converged extremely slowly), or either because none exist or they exist in a narrow band in space and period length making it difficult to find numerically.

    In Figure~\ref{fig:cluster} (right-hand-side) we show the projection of the UPOs, on the two variables of interest,  that we found on the attractor. This shows how the UPOs appear to form dense sets over the dense areas in the attractor. It is expected that such numerical searches for UPOs will be biased towards short period UPOs, as long period UPOs are numerically more difficult to find~\citep{cvitanovic2020}. In our case, however, we have found approximately an equal number of UPOs between periods of 4-45 days. This is clearly not a complete set of UPOs for these periods, and the skew in the period of the resulting UPOs found is likely due to the most common repeating cycles in the key variables ($\QgsGroundTemp[3],~\QgsBarotropic[2]$). This is not unexpected, as the numerical method we employ is highly sensitive to the initial conditions used. In our case, the initial conditions were primarily found by looking for close recurrences in the trajectory, which in turn is driven by natural cycles in these variables.

\subsection{Origin of Blocking Clusters}\label{sec:results_sub:cluster_origin}
\begin{figure}
        \centering
        \includegraphics[width=\linewidth]{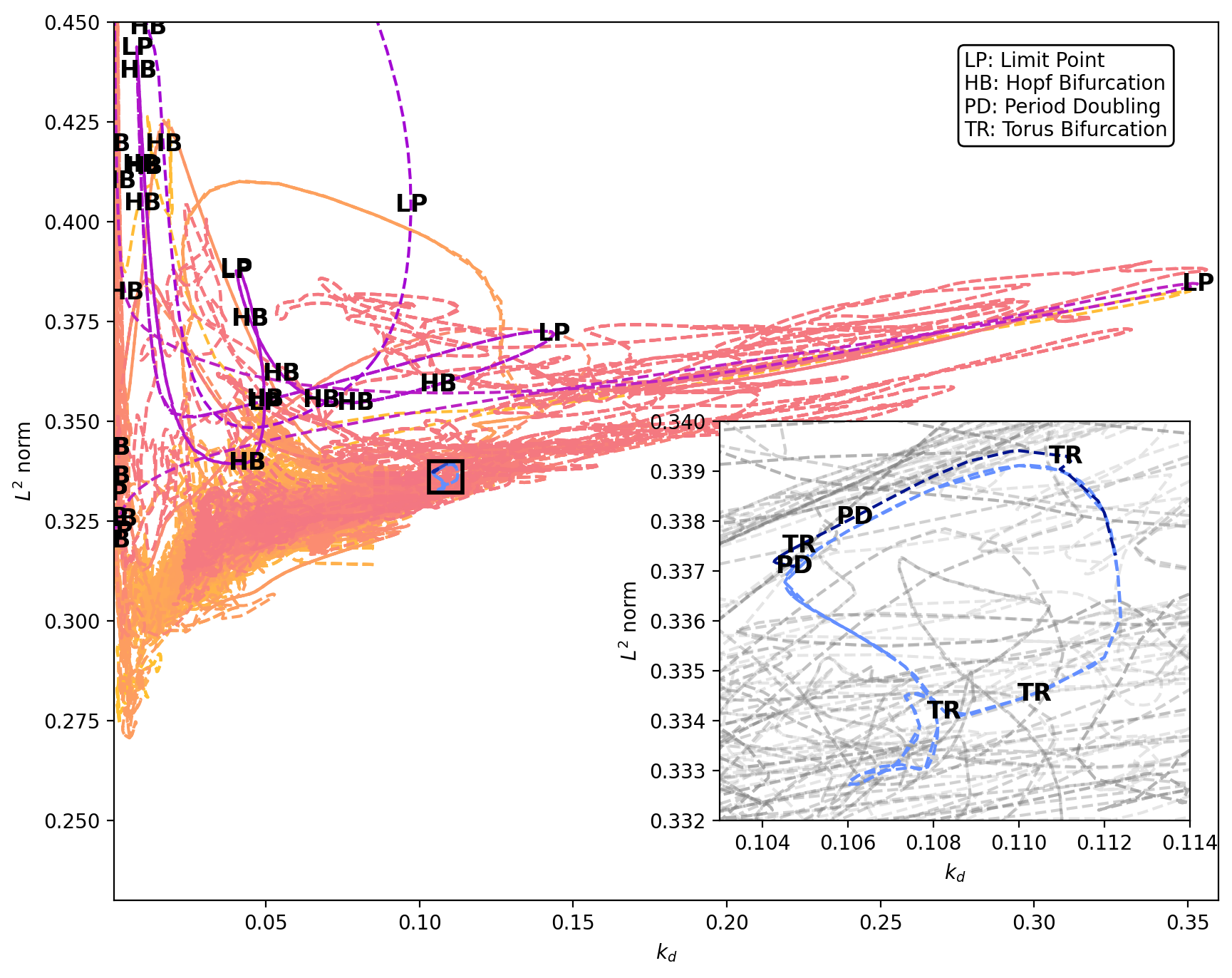}
            \caption{Bifurcation diagram of all fixed points (shown in purple) and the branching periodic orbits (shown in yellow, orange, and pink to create some visual distinction) of the model for varying $\QgsFriction$ with a value of $\QgsAtmInsolation=300\si{\watt\per\square\meter}$. Dotted lines represent unstable regions and solid lines represent stable regions. Bifurcation points are shown for the fixed points continuations. The black rectangle is zoomed in on and shown in the inset, and focuses on two key branches shown in blue. These two periodic branches were found by continuing periodic orbits that have a small window of stability; The light blue branch being stable at $\QgsFriction\approx0.112$ and the dark blue at $\QgsFriction\approx0.1045$.}
            \label{fig:k_d_bif_diag}
    \end{figure}

\begin{figure}
        \centering
        \includegraphics[width=\linewidth]{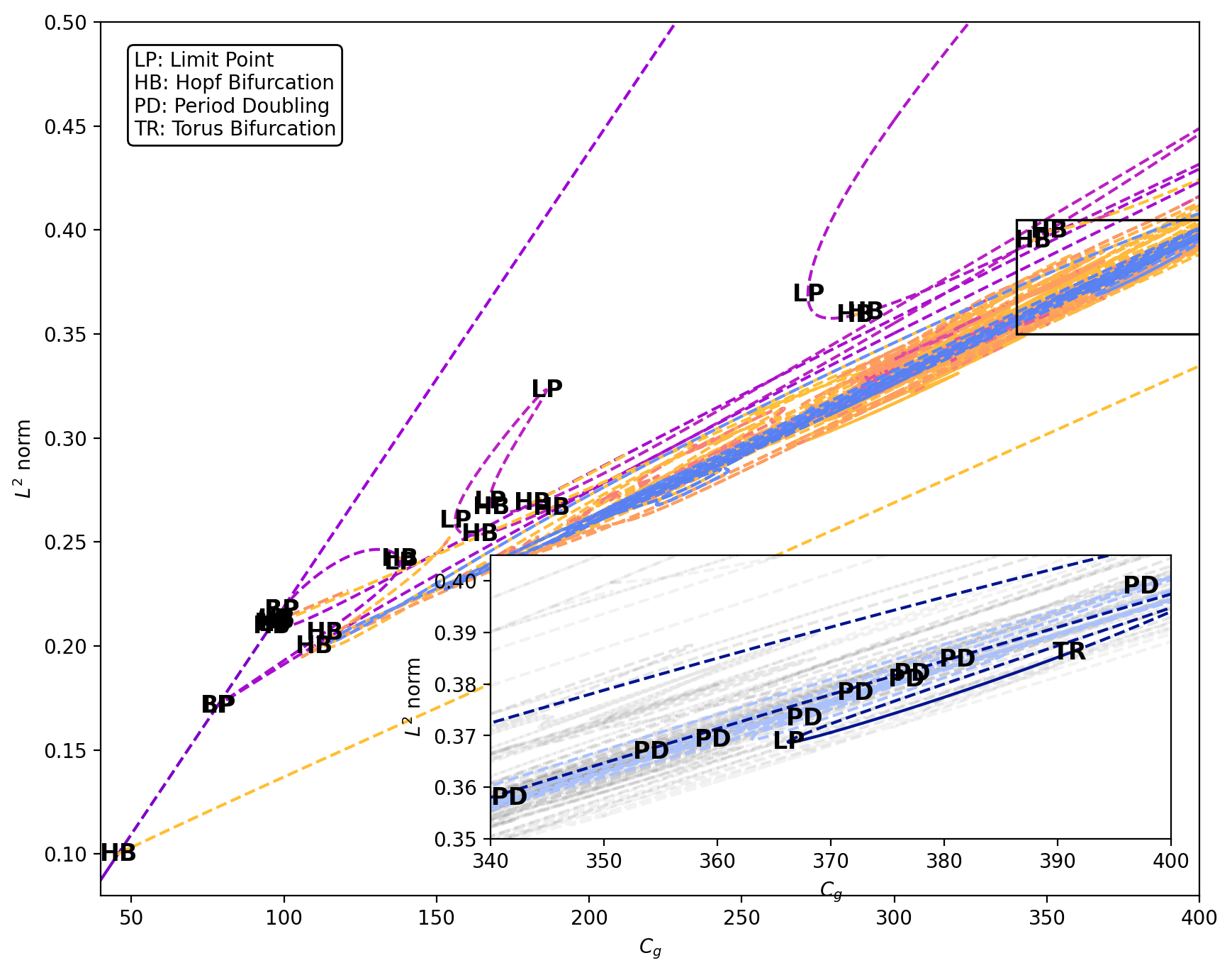}
        \caption{Bifurcation diagram for all fixed points (shown in purple) and the bifurcation and branching points that lead to the periodic orbits (shown in yellow, orange, and pink to create some visual distinction) of the model for varying $\QgsAtmInsolation$ and fixed $\QgsFriction=0.085$. The dotted lines represent unstable regions and the solid lines represent stable regions. The black rectangle is zoomed in on and shown in the inset, and focuses on four key branches of periodic orbits, shown in blue. A stable periodic orbit is found for large values of $\QgsAtmInsolation\approx380\si{\watt\per\square\meter}$ (shown in dark blue), stability is lost through a fold and a torus bifurcation. This branch is tracked along with any branching points or period doubling branches (these are shown in light blue).}
        \label{fig:c_g_bif_diag}
    \end{figure}

    To understand why our model contains these two regimes, we undertook a bifurcation analysis of the model for varying $\QgsFriction$ and $\QgsInsolation$, as varying these parameters resulted in the finding of stable periodic solutions by~\citet{xavier2024}. We have developed on this previous work by using AUTO~\citep{doedel2012} to investigate the bifurcation structure that originates from these two stable windows. Figure~\ref{fig:k_d_bif_diag} shows the bifurcation diagram as a function of $\QgsFriction$ and Figure~\ref{fig:c_g_bif_diag} shows that of $\QgsInsolation$. These bifurcation diagrams were produced by first numerically finding fixed points of the model and continuing along all branches, and following any branching points. Once the fixed points continuations were produced, we branched at any detected Hopf bifurcation points, and the corresponding periodic orbit branches were followed, again branches originating from periodic orbit branching points or period doubling bifurcations were also followed. This process was automated using the auto-AUTO package. Lastly, we also continued the periodic orbit branches, from the UPOs found numerically using the Newton-Raphson method described in Section~\ref{sec:methods_sub:finding_upos} and included these in the bifurcation diagram. 

    There exists a window, at approximately $\QgsFriction\approx0.1045$, and $\QgsFriction\approx0.115$, where periodic orbits living in the LHS cluster becomes stable. We used AUTO to track these orbits to investigate how they destabilise and whether they can give a hint at how chaos emerges. In Figure~\ref{fig:k_d_bif_diag} we show these regions of stability on the bifurcation diagram. We see that there is a closed branch (shown in light blue) which goes through a period doubling bifurcation, and leads to another branch (shown in dark blue) that has a small window of stability before destabilising through a torus bifurcation. The bifurcation diagram shows that this structure has a similar value in $\lNorm$ norm space to the other periodic orbit branches. This means that while we cannot see how these key branches connect to the rest of the bifurcation diagram, we hypothesise that these branches originate from a torus bifurcation of another branch, a guess that is reinforced by the visual similarity between these orbits and other orbits that we found in this cluster.

    \begin{figure}
        \centering
        \begin{subfigure}{\textwidth}
            \centering
            \includegraphics[width=\linewidth]{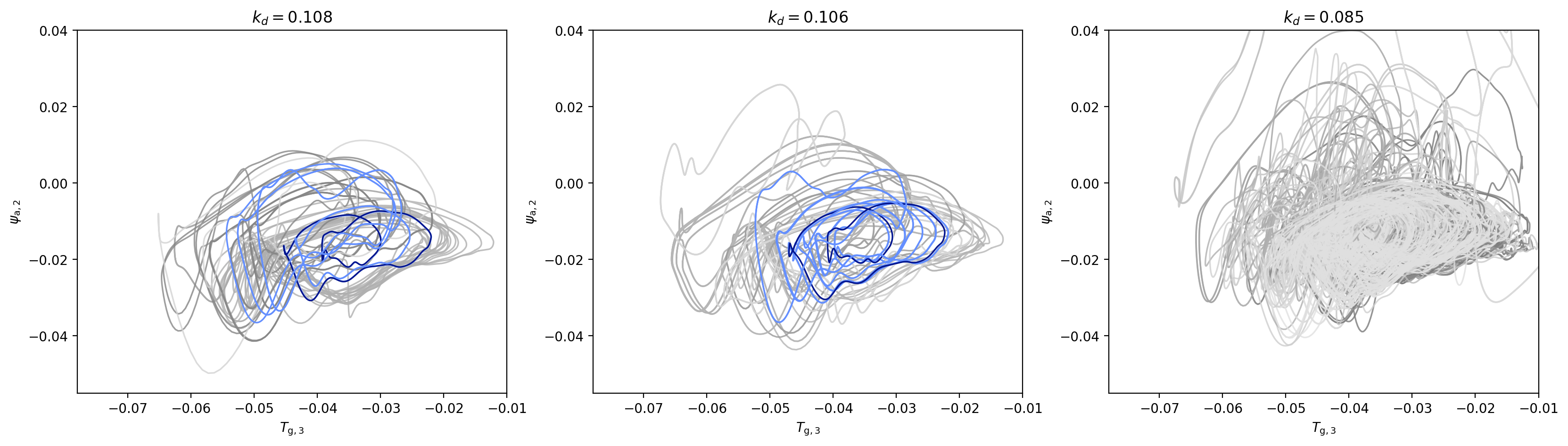}
            \caption{}
            \label{fig:Cluster_origin_sub:LHS}
        \end{subfigure}
        \vspace{0.2cm}
        \begin{subfigure}{\textwidth}
            \centering
            \includegraphics[width=\linewidth]{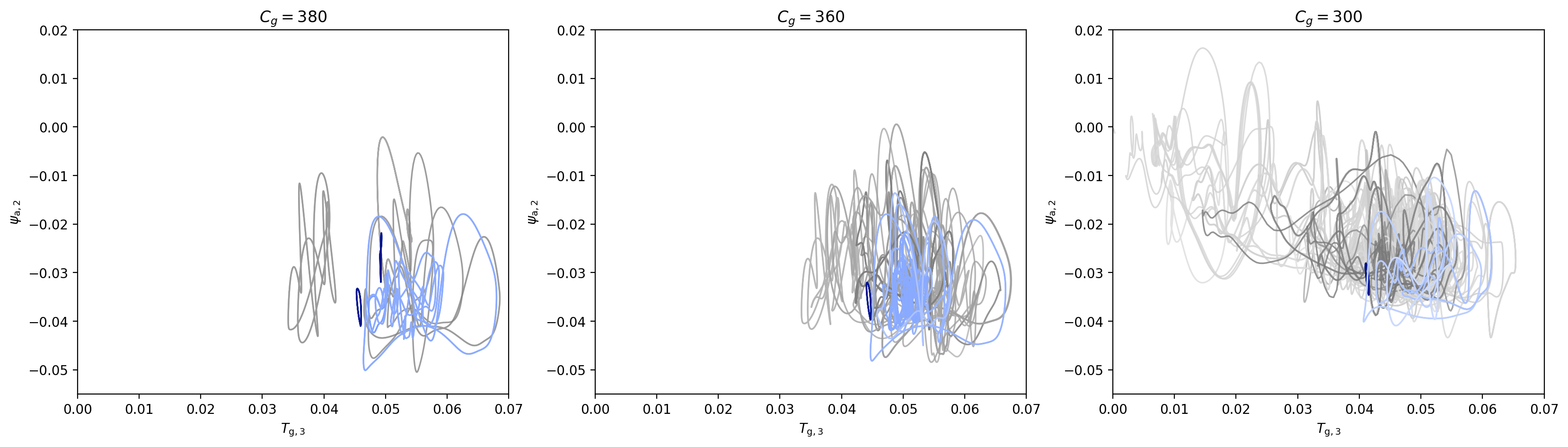}
            \caption{}
            \label{fig:Cluster_origin_sub:RHS}
        \end{subfigure}
        \caption{State space for variables ($\QgsGroundTemp[3],~\QgsBarotropic[2]$), for decreasing values of $\QgsFriction$ (Figure~\ref{fig:Cluster_origin_sub:LHS}) and $\QgsInsolation$ (Figure~\ref{fig:Cluster_origin_sub:RHS}). The orbits shown in blues correspond to the branches shown in the inset plots in Figures~\ref{fig:k_d_bif_diag} and~\ref{fig:c_g_bif_diag} respectively. Grey orbits are show the other branches found, that do not directly branch from the key orbits. On the left we show the parameter values where a stable orbit was found, we then decrease the parameter values to end up with the clusters, as shown in Figure~\ref{fig:cluster}.}
        \label{fig:Cluster_origin}
    \end{figure}

    For parameter values around $\QgsFriction=0.085$ and $\QgsInsolation=380\si{\watt\per\square\meter}$, there exists a stable periodic orbit, centred in the right hand cluster. We tracked this orbit for $\QgsInsolation$ (Figure~\ref{fig:c_g_bif_diag}). For decreasing $\QgsInsolation$ a fold bifurcation occurs and the orbit looses stability, and for increasing the parameter the orbit looses stability at a torus bifurcation. We see that this branch has a number of period doubling bifurcations, and following these bifurcations we see that a large number of periodic branches stem from this original branch.

    The origins of each cluster is displayed in Figure~\ref{fig:Cluster_origin}. In this image we show the left hand cluster (Figure~\ref{fig:Cluster_origin_sub:LHS}), for decreasing values of $\QgsFriction$. Two particular UPOs are shown, coloured to match the colouring in the bifurcation diagram shown in Figure~\ref{fig:k_d_bif_diag}. All other periodic orbits are shown in grey. The blue UPOs appear to occupy the central section of the cluster defined by the UPOs, and we see that this structure remains even once the coloured periodic orbit branches go through a fold bifurcation and disappear for lower values of $\QgsFriction$. For the right hand side cluster (Figure~\ref{fig:Cluster_origin_sub:RHS}) we decrease the value of $\QgsInsolation$,and present UPOs coloured as in the inset in Figure~\ref{fig:c_g_bif_diag}. Here we see that they again form a central structure within the cluster of UPOs, and for this parameter values we can follow the branches to the values of $\QgsInsolation$ that we use in the analysis. Together these key branches provide a hint at the origin of the cluster dynamics that we find in this model.

\subsection{Clustering the Attractor}\label{sec:results_sub:clustering}
    In this study we use the UPOs to assign each point of a trajectory into the left and right hand clusters identified, and shown on Figure \ref{fig:cluster}. This is done by first splitting the UPOs into two clusters dependent on if they have positive or negative values of $\QgsGroundTemp[3]$. More precisely, we analyse the maximum and minimum values of each UPO on the variable $\QgsGroundTemp[3]$, which is the ground heat projected onto the mode $F_3=2\sin(nx)\sin(y)$:
    \begin{align*}
        \text{LHS Cluster UPOs}:\ C^{\text{UPO}}_{\text{LHS}} &= \{\UPO{\UPOperiod}{i}:\ \max_{\QgsGroundTemp[3]}{\UPO{\UPOperiod}{i}} < 0, \ i\in I\}\\
        \text{RHS Cluster UPOs}:\ C^{\text{UPO}}_{\text{RHS}} &= \{\UPO{\UPOperiod}{i}:\ \min_{\QgsGroundTemp[3]}{\UPO{\UPOperiod}{i}} > 0, \ i\in I\}
    \end{align*}
    where $\UPO{\UPOperiod}{i}$ is a UPO with a period $\UPOperiod$, and an index $i$, and $I$ is an indexing of all UPOs found. This results in 3425 UPOs in the left hand cluster, and 3361 on the right (32 UPOs do not fall into either cluster and are ignored). Next, for a given trajectory $\systemFlow(\IC[j])$ with initial condition $\IC[j]$, we cluster the points of the trajectory ($\systemFlow[t_k](\IC[j])$) by calculating which UPO the given point is closest to, and then assign the cluster from which cluster the UPO is a member of:
    \begin{align*}
        \text{LHS Cluster Trajectory}:\ C^{\systemFlow}_{\text{LHS}} &= \{\systemFlow[t_k](\IC[j]): \min_{i\in I}\left\|\systemFlow[t_k](\IC[j])- \UPO{\UPOperiod}{i}\right\|,\ \UPO{\UPOperiod}{i}\in C^{\text{UPO}}_{\text{LHS}}\}\\
        \text{RHS Cluster Trajectory}:\ C^{\systemFlow}_{\text{RHS}} &= \{\systemFlow[t_k](\IC[j]): \min_{i\in I}\left\|\systemFlow[t_k](\IC[j])- \UPO{\UPOperiod}{i}\right\|,\ \UPO{\UPOperiod}{i}\in C^{\text{UPO}}_{\text{RHS}}\}
    \end{align*}
    where the notation $\left\|\cdot\right\|$ stands for the Euclidean norm.

    The clustering is visualised by the color scheme (pink and orange) in Figure~\ref{fig:cluster}. As we see, this is not as simple as just clustering the points of the trajectory by value of $\QgsGroundTemp[3]$ as in other projections the UPOs are not neatly separated. We think this clustering method is optimal as the UPOs form a dense skeleton in dense region of the attractor, and by clustering on which UPO a trajectory point is closest to, it is expected that the UPO should provide some description as to the flow in that neighbourhood. Of course this approach is dependent on the number of UPOs found and how close the trajectory is to a given UPO. If few UPOs are found, the attractor is not described in any real detail and if the trajectory is far from a UPO then there is no reason to believe that the closest UPO will have any influence on the trajectory.

    We expect that our approach is robust to the above concerns. It has been found in that a surprising low number of UPOs still provides reasonable results~\citep{hunt1996}. In our case there is clearly multiple distinct regimes within the single attractor~\citep{xavier2024}, and the clusters found here overlap well with those found in this earlier study. We believe that our clustering method provides a cleaner method for isolating transitions as it is not a probabilistic clustering method. In addition, it provides a measure between a point of the trajectory and the cluster of UPOs that covers the dense regions of the attractor. This distance therefore provides a useful metric for identifying transitions, as well as providing information on how much we expect the cluster of UPOs to effect the behaviour of a trajectory.

\subsection{Transition paths}\label{sec:results_sub:transitions}
\begin{figure}
        \centering
        \begin{subfigure}{0.8\textwidth}
            \centering
            \includegraphics[width=\linewidth]{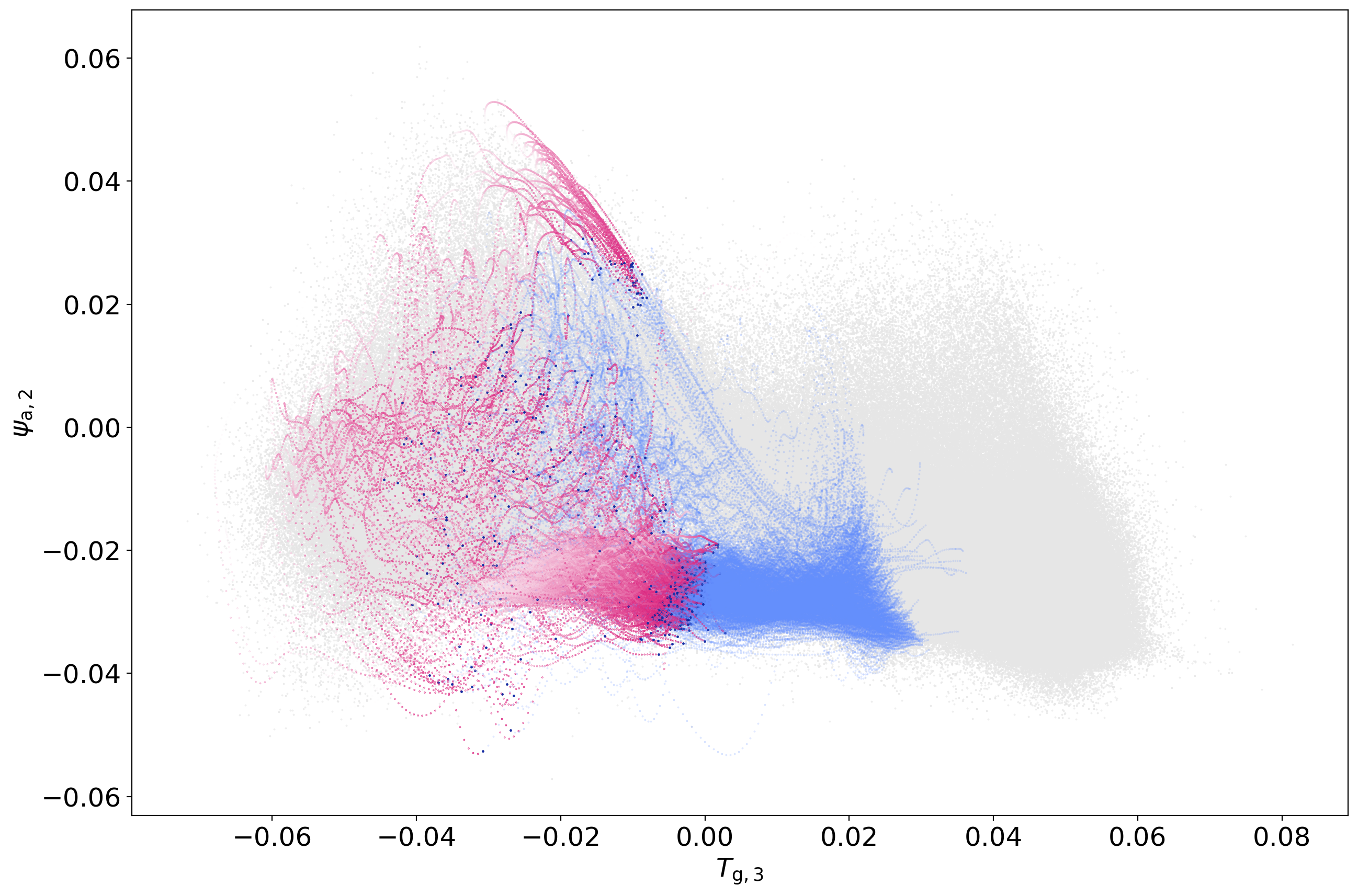}
            \caption{}
            \label{fig:lower_transition_sub:introduction}
        \end{subfigure}
        \begin{subfigure}{0.8\textwidth}
            \centering
            \includegraphics[width=\linewidth]{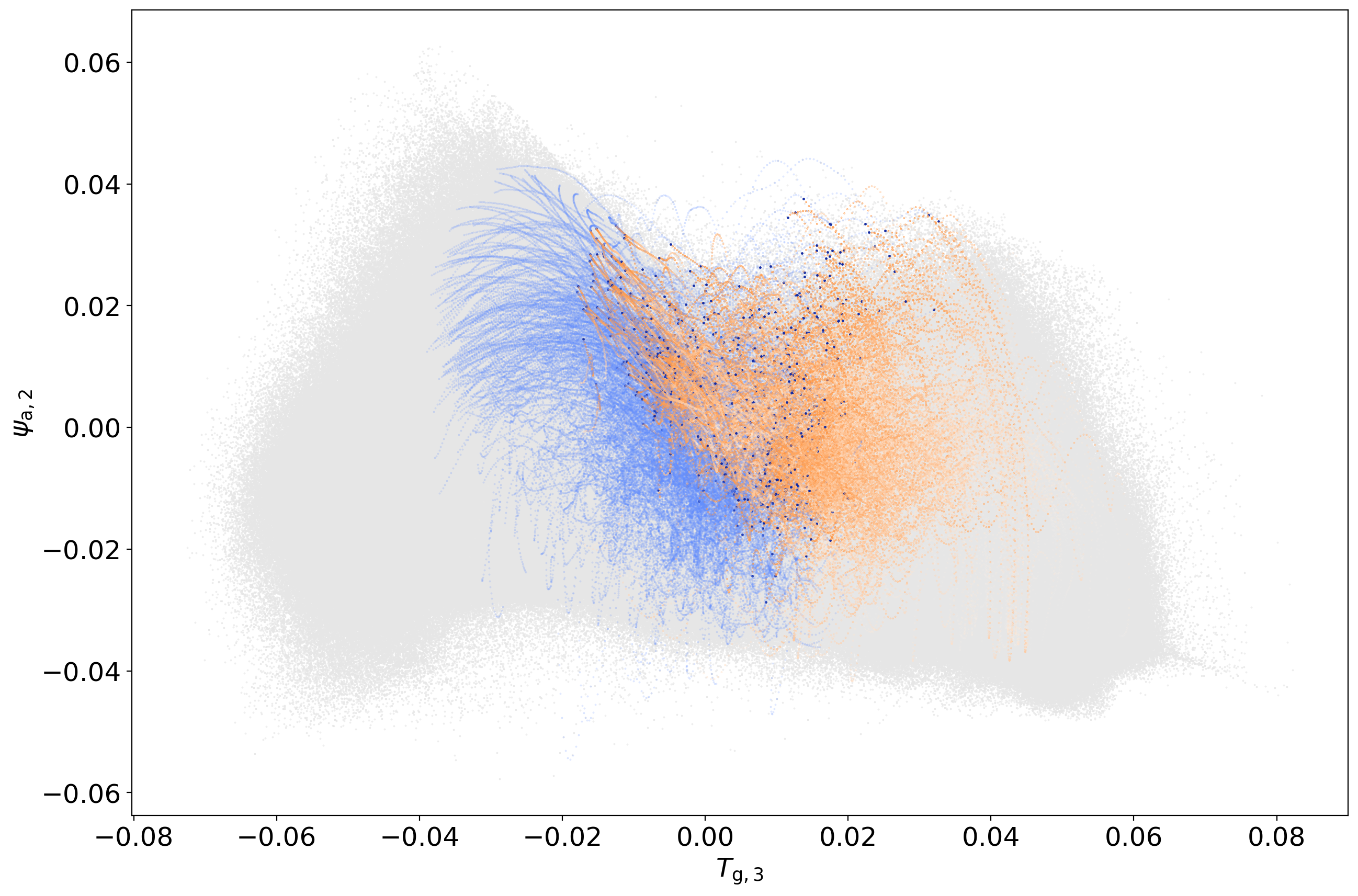}
            \caption{}
            \label{fig:r_to_l_transition_sub:introduction}
        \end{subfigure}
        \caption{Transitions from one cluster to the other, Figure~\ref{fig:lower_transition_sub:introduction} shows transitions from left to right, and Figure~\ref{fig:r_to_l_transition_sub:introduction} from right to left. The trajectory is shown 11 days before the transition (in pink or orange, where the saturation increases towards the transition). The transition points are shown in navy blue, and the post transition trajectory is shown in light blue for 11 days after the transition.}
    \end{figure}

    Transitions between the two regimes of the attractor occur through specific regions of the state space, shown in Figures~\ref{fig:lower_transition_sub:introduction} and~\ref{fig:r_to_l_transition_sub:introduction}. The transitions are identified by the points in state space where a trajectory passes from one cluster to another, and by analysing the points on the trajectory that precede and following the transition. To analyse the properties of these transitions we began by running a long trajectory, and then isolating the occurrences where a transition occurs. To ensure we are capturing transitions between the regimes we filter the points to ensure that a trajectory remained in one cluster for at least 56 days before, and 56 days in the other cluster after transitioning. This threshold is set to remove points that are in between the two sets of UPOs and are therefore not described well by our clustering method (approximately 0.07\% of points).

    \begin{figure}
        \centering
        \begin{subfigure}{0.9\textwidth}
            \centering
            \includegraphics[width=\linewidth]{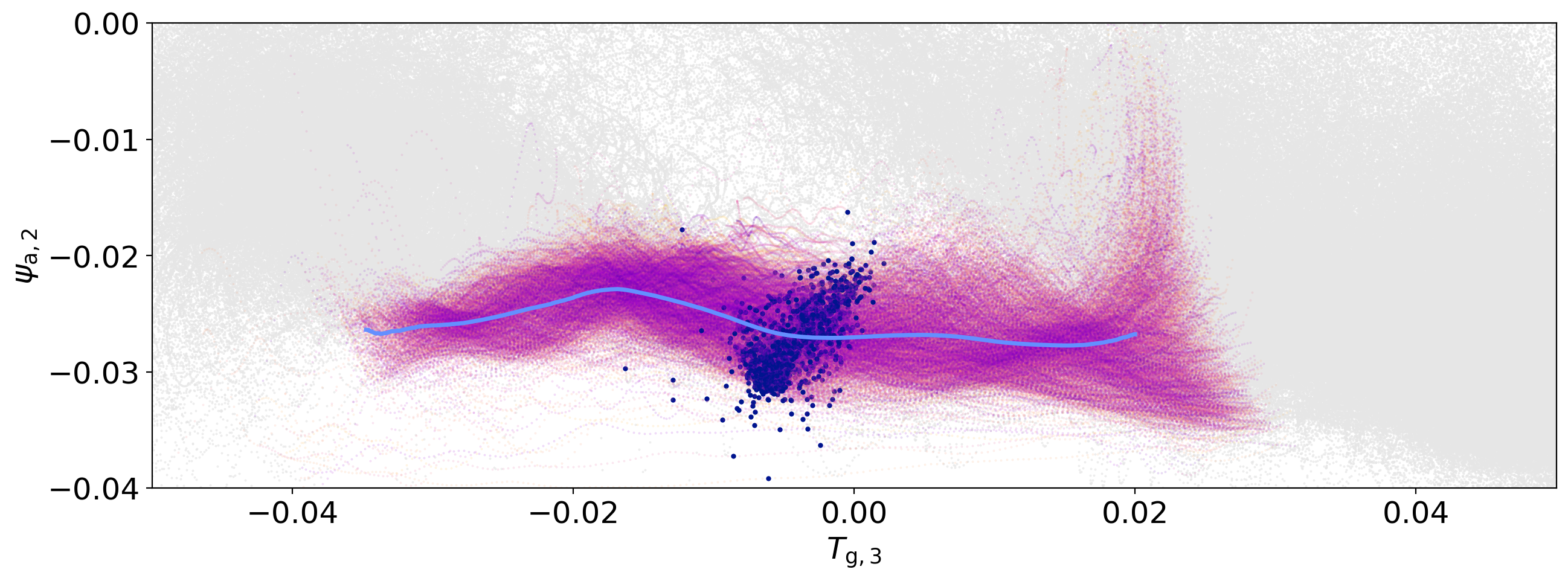}
            \caption{}
            \label{fig:lower_transition_sub:key_transitions}
        \end{subfigure}
        \vspace{0.2cm}
        \begin{subfigure}{0.9\textwidth}
            \centering
            \includegraphics[width=\linewidth]{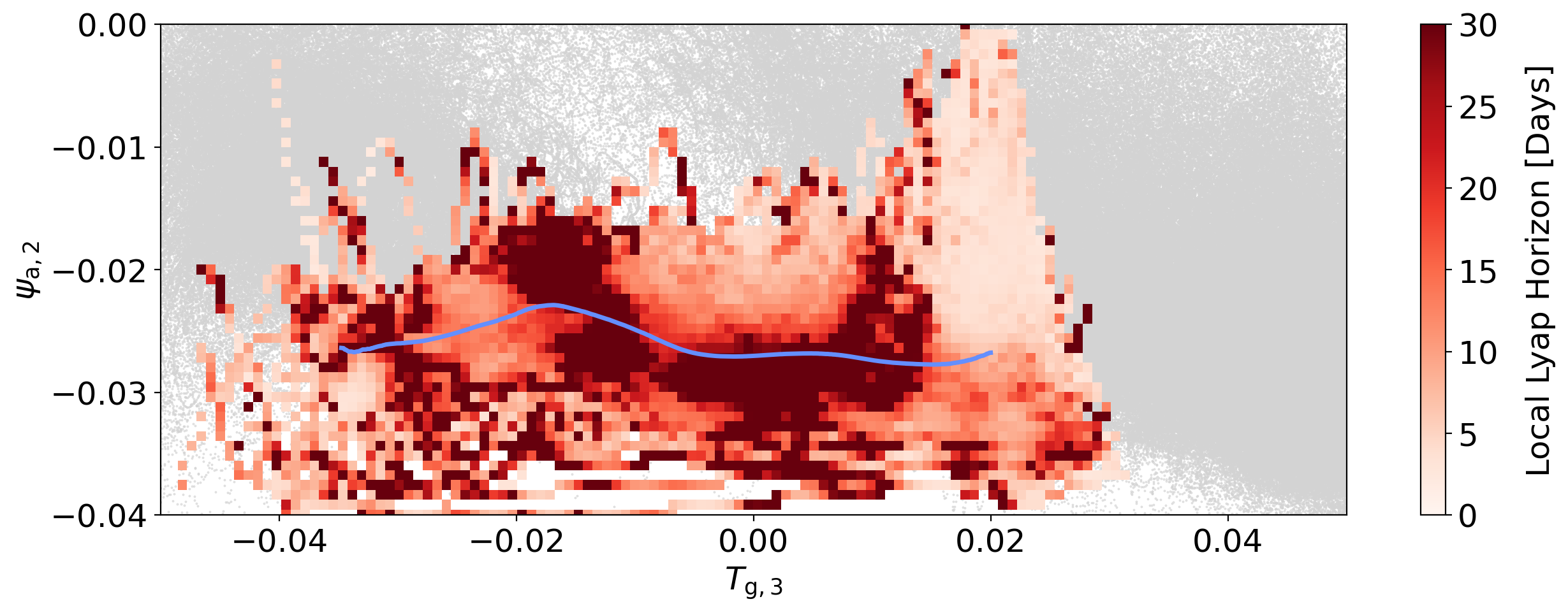}
            \caption{}
            \label{fig:lower_transition_sub:LLE_horizen}
        \end{subfigure}
        \caption{Left to right transitions, taking place in the lower section of the attractor. Figure~\ref{fig:lower_transition_sub:key_transitions} shows the filtered trajectories that transition from left to right in the lower portion of the attractor. The mean of these trajectories is shown in light blue. The transition points are shown in navy blue. Figure~\ref{fig:lower_transition_sub:LLE_horizen} displays local Lyapunov horizon, calculated by taking the inverse of the binned local Lyapunov exponent.}
        \label{fig:lower_transition}
    \end{figure}

    \begin{figure}
        \centering
        \begin{subfigure}{0.9\textwidth}
            \centering
            \includegraphics[width=\linewidth]{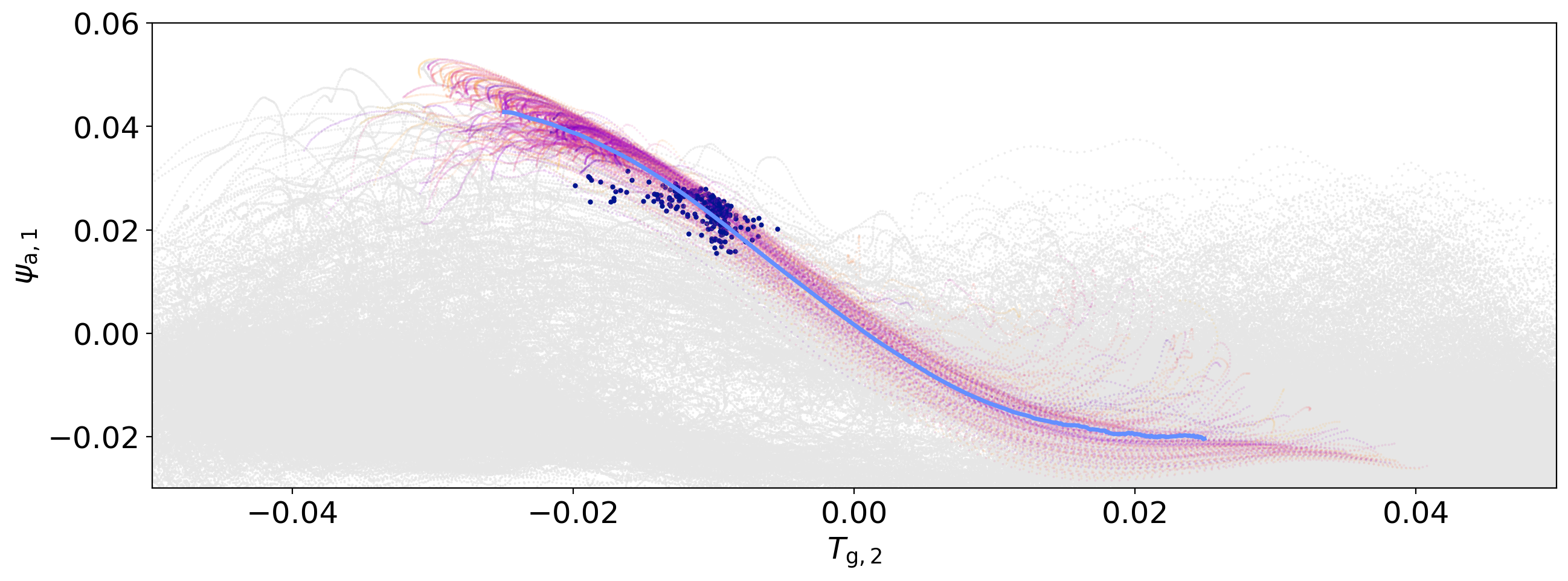}
            \caption{}
            \label{fig:upper_transition_sub:key_transitions}
        \end{subfigure}
        \vspace{0.2cm}
        \begin{subfigure}{0.9\textwidth}
            \centering
            \includegraphics[width=\linewidth]{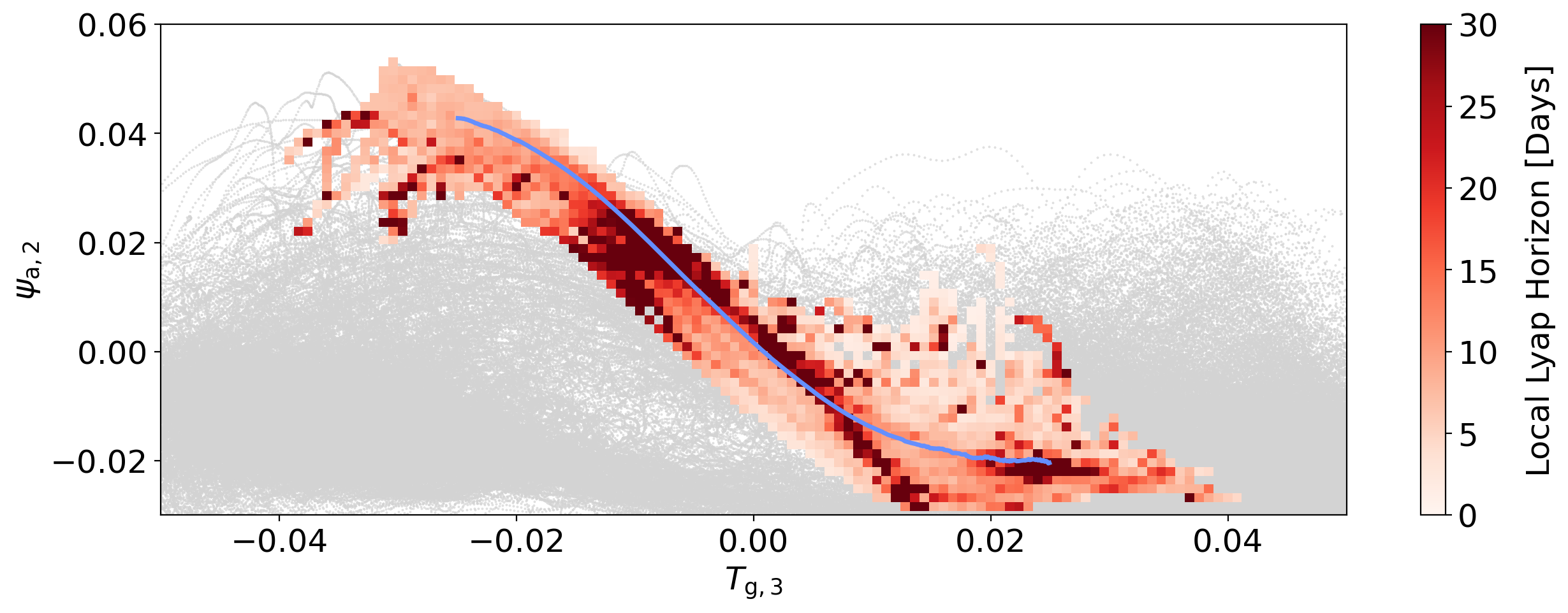}
            \caption{}
            \label{fig:upper_transition_sub:LLE_horizen}
        \end{subfigure}
        \caption{Upper left to right transition. Figure~\ref{fig:upper_transition_sub:key_transitions} shows the filtered trajectories that transition from left to right in the upper section of the attractor, shown in Figure~\ref{fig:lower_transition_sub:introduction}, with the same colour scheme. Figure~\ref{fig:upper_transition_sub:LLE_horizen} shows the same as Figure~\ref{fig:lower_transition_sub:LLE_horizen}, but again for the upper left to right transition.}
        \label{fig:upper_transition}
    \end{figure}

\begin{figure}
        \centering
        \begin{subfigure}{\textwidth}
            \centering
            \includegraphics[width=\textwidth]{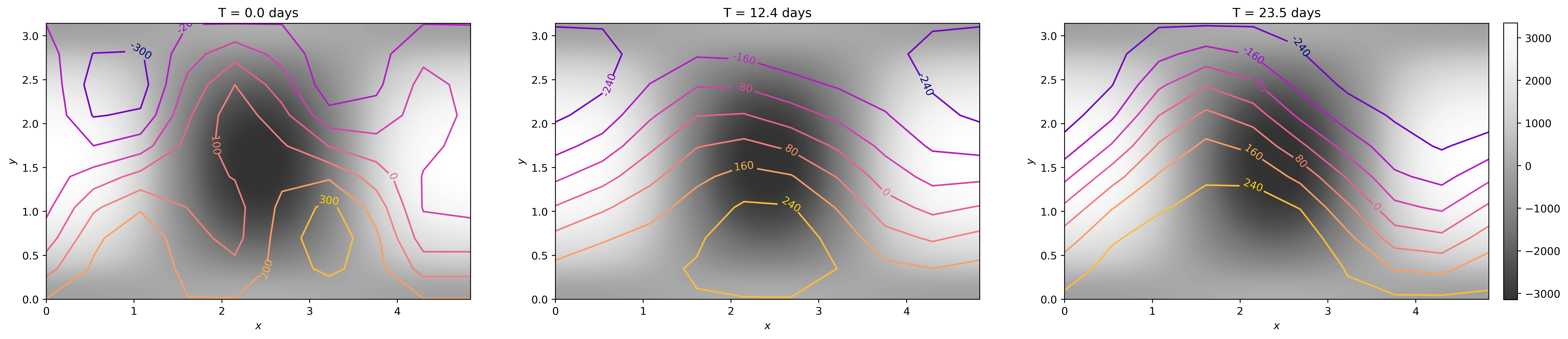}
            \caption{Transition from left to right in the lower portion of the attractor, shown in Figure~\ref{fig:lower_transition_sub:key_transitions}. Here the average blocking behaviour transitions from the windward side of the orography to the leeward side by travelling to the west.}
            \label{fig:lower_climate}
        \end{subfigure}
        \vspace{0.5cm}
        \begin{subfigure}{\textwidth}
            \centering
            \includegraphics[width=\textwidth]{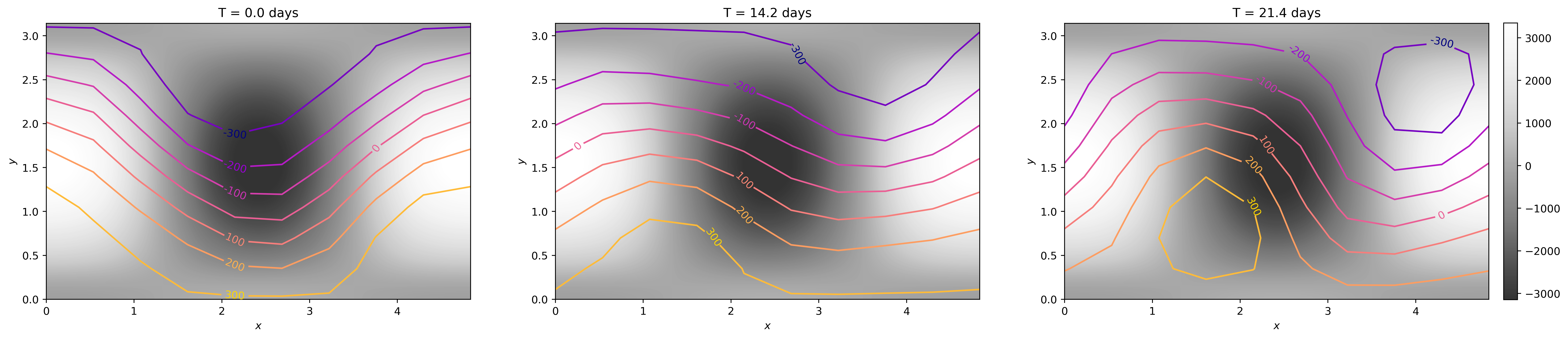}
            \caption{Transition from left to right in the upper portion of the attractor, shown in Figure~\ref{fig:upper_transition}. The blocking transitions from above the orography to the leeward side by travelling to the east.}
            \label{fig:upper_climate}
        \end{subfigure}
        \vspace{0.5cm}
        \begin{subfigure}{\textwidth}
            \centering
            \includegraphics[width=\textwidth]{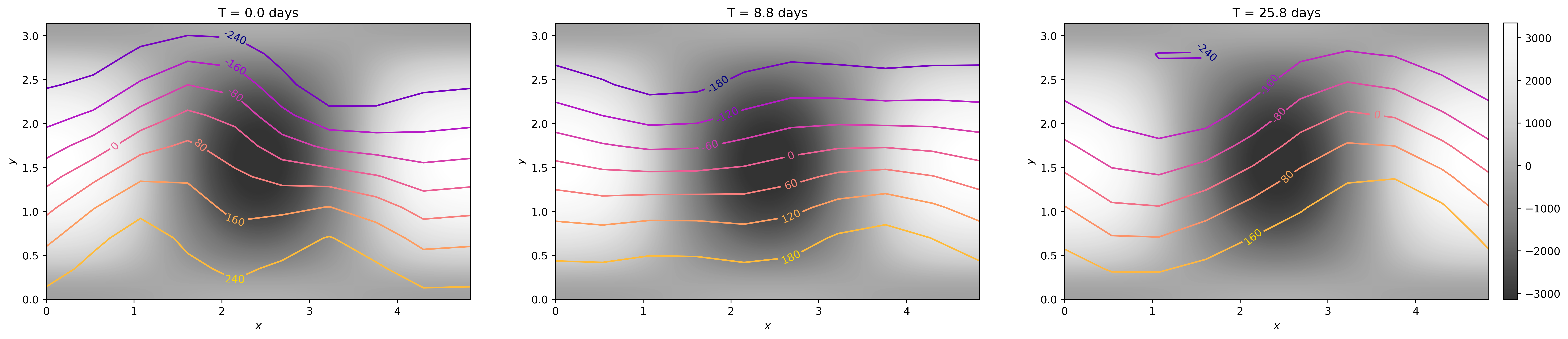}
            \caption{Transition from right to left, as in Figure~\ref{fig:r_to_l_transition_sub:key_transitions}. Here the blocking moves from the leeward side of the orography to the windward side through zonal atmospheric behaviour.}
            \label{fig:r_to_l_climate}
        \end{subfigure}
        \caption{Snap shots of the climatology of the averaged transition trajectories. The model time of each snap shot is shown in the title. The contours show the $500\si{\hecto\pascal}$ height, and the greyscale shading shows the relative height of the orography in metres.}
        \label{fig:climate}
    \end{figure}

    By parametrising on $\QgsGroundTemp[3]$ we created averaged transition trajectories to analyse the climatology of these paths. Transitions from the LHS to RHS cluster appear to pass through two regions, and the averaged trajectories shown are the transitions between two blocking regimes. This is primarily driven by the ground temperature anomaly shifting between the east and west. The transition in the lower portion of the attractor, shown in Figure~\ref{fig:lower_transition_sub:key_transitions} displays the blocking regime moving to the west, which can be seen in the model climatology shown in Figure~\ref{fig:lower_climate}. The other left to right transition taking place in the upper part of the attractor (Figure~\ref{fig:upper_transition_sub:key_transitions}) shows the blocking regime moving to the east (Figure~\ref{fig:upper_climate}).

\begin{figure}
        \begin{subfigure}{0.8\textwidth}
            \centering
            \includegraphics[width=\linewidth]{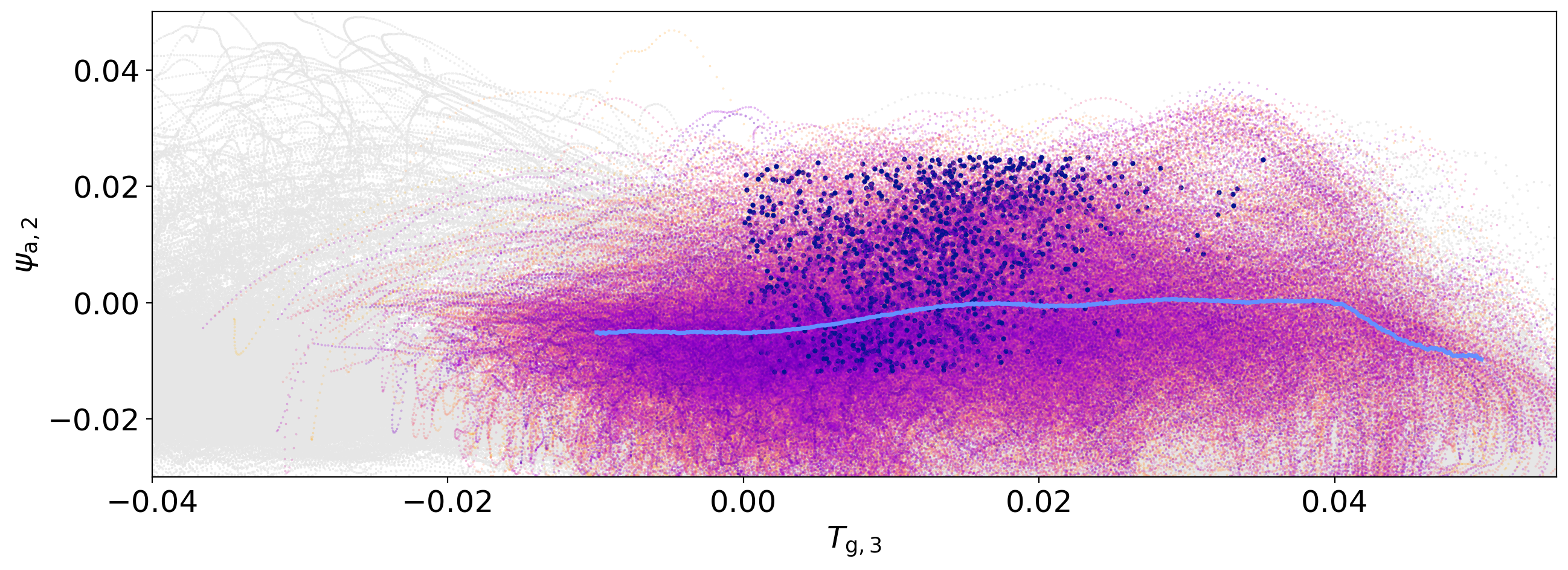}
            \caption{}
            \label{fig:r_to_l_transition_sub:key_transitions}
        \end{subfigure}

        \vspace{0.2cm}

        \begin{subfigure}{0.8\textwidth}
            \centering
            \includegraphics[width=\linewidth]{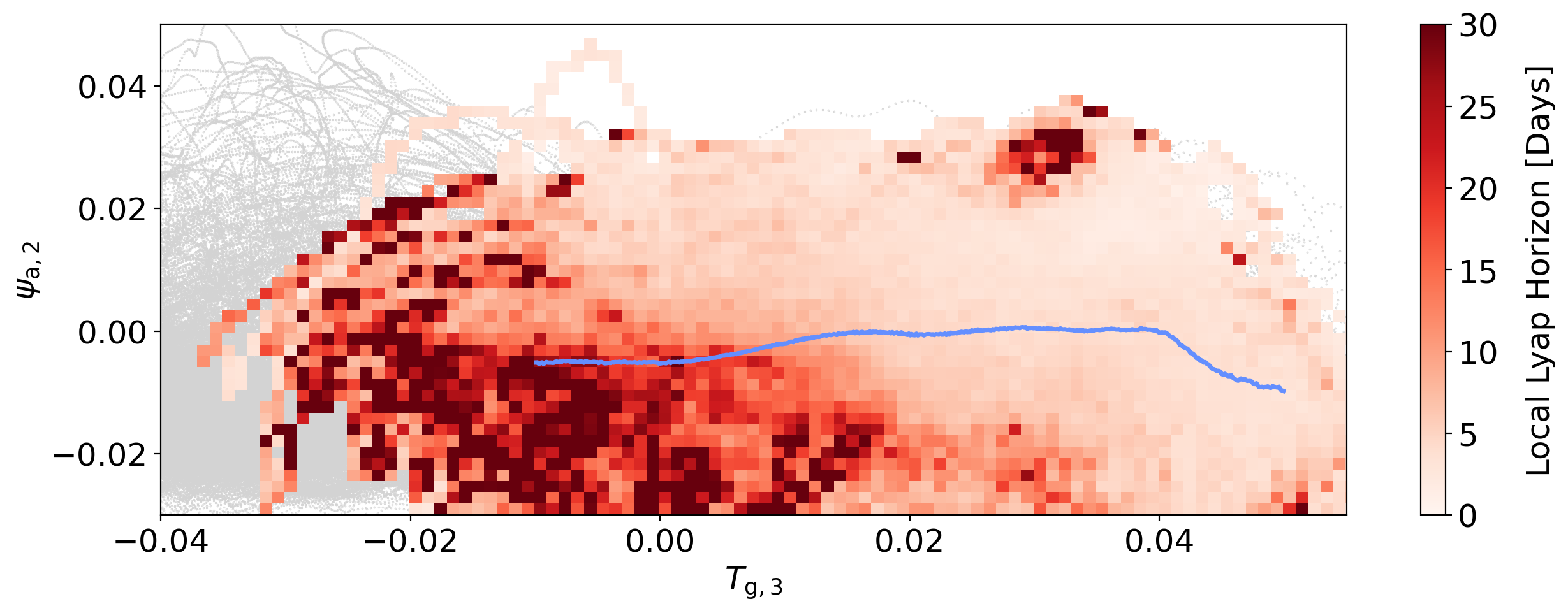}
            \caption{}
            \label{fig:r_to_l_transition_sub:LLE_horizen}
        \end{subfigure}
        \caption{Right to left transition. Figure~\ref{fig:r_to_l_transition_sub:key_transitions} shows filtered transition trajectories. The mean of these trajectories is shown in light blue. The transition points are shown in navy blue. Figure~\ref{fig:r_to_l_transition_sub:LLE_horizen} displays the same as Figure~\ref{fig:lower_transition_sub:LLE_horizen}, but for the right to left transition.}
        \label{fig:right_to_left_transition}
    \end{figure}

    Transitions from the RHS to LHS are not as clearly defined in state space (shown in Figure~\ref{fig:r_to_l_transition_sub:introduction}), but the averaged transition (Figure~\ref{fig:r_to_l_transition_sub:key_transitions}) show that the transitions between the two blocking regimes result in averaged zonal behaviour in the atmosphere. This is not surprising as the transitions paths pass close to the Hadley circulation. In~\citet{xavier2024} the zonal regime was identified as a separate cluster, but in our analysis we find that this atmospheric regime primarily occurs during transitions between the RHS cluster to the LHS cluster, as shown in Figure~\ref{fig:r_to_l_climate}.

    Finally, 55 percent of transitions between regimes do not occur through the regions described above. These transitions occur for trajectories that spend short time in a given regime (less than 56 days). These transitions do not appear to follow predictable behaviour and are probably due to the trajectory making diversions in other dimensions and the transitions only appear to occur due to the projection we make onto two dimensions.

    The paths of the transitions were found to occur through relatively stable regions of the state space. This is shown in Figures~\ref{fig:lower_transition_sub:LLE_horizen}, \ref{fig:upper_transition_sub:LLE_horizen}, \ref{fig:r_to_l_transition_sub:LLE_horizen} which show the Lyapunov horizon, calculated by taking the local Lyapunov exponent for each transition, binning the results in state space, and then taking the inverse of the results. This shows that while the onset of the transition occurs through an unstable region, once the trajectory begins transitioning from one regime to another, the path it takes, and therefore the atmospheric behaviour, is more predictable. 

\subsection{UPO Shadowing}\label{sec:results_sub:upo_shaddowing}
    The trajectory moves between the two clusters, sometimes spending a long time in one regime (up to 1000 days in the RHS cluster and 600 days in the LHS), other times quickly returning back to the previous regime. Above, we showed that when a trajectory spends long times in one regime and then transitions to the other regime, it tends to pass through a well-defined region in the state space. We now look at which collection of UPOs the trajectory passes close to during its visit to a cluster. The idea being that if the transition path appears to be confined in state space, the trajectory must pass close to a specific set of UPOs while transitioning.  As UPOs are dense in the attractor, the UPOs can provide an arbitrarily accurate approximation for a given trajectory~\citep{maiocchi2022}. The idea of looking at UPOs close to the trajectory, to gain some understanding of the future path the trajectory will take, is referred to as UPO shadowing. In our case, we have only found a small selection of UPOs so the accuracy of any shadowing will be limited, but the idea remains the same; a UPO will shadow a trajectory if they are close to one another, and for some limited time the evolution of the trajectory is similar to that of the UPO.

    More concretely, a UPO $\UPO{\UPOperiod}{i}$ shadows the trajectory at time $t$ if it is the closest UPO, of all UPOs found, to the location of the trajectory at time $t$: $\min_{i\in I}\left\|\UPO{\UPOperiod}{i} - \systemFlow[t](\IC[j])\right\|$, where $I$ is an index of all the UPOs. Here we are assuming that we can create an index of the UPOs as numerically we can only find a finite number of UPOs, the method of indexing the UPOs is arbitrary. We are interested in which UPO is the closest at a given time, so we define a function $\UPOIndexFunc(\systemFlow[t](\IC[j]))$ which returns the index of the UPO, given the trajectory point $\systemFlow[t](\IC[j])$, where $\UPOIndexFunc:\mathbb{R}^n\rightarrow I$, and $I$ is the indexing set of the UPOs.

    We define the number of times that a given UPO shadows a trajectory over a set period of time, which we call cumulative shadowing (CS). This is done by counting the number of times a given UPO $\UPO{\UPOperiod}{i}$ shadowed the trajectory in the time range $(t_1,~t_2)$, denoted as $\shadowing(\UPO{\UPOperiod}{i},~t_1, ~t_2)$:
    
    \begin{equation}
        \shadowing(\UPO{\UPOperiod}{i},~t_1, ~t_2) = \sum_{t\in[t_1,~t_2]} \delta\bigg(i,~\UPOIndexFunc(\systemFlow[t](\IC[0])) \bigg) \vspace{0.5cm}
    \end{equation}

    where $\delta$ is the Kronecker delta, and returns 1 if the closest UPO has index $i$, otherwise returns 0.

\begin{figure}
        \centering
        \includegraphics[width=0.9\linewidth]{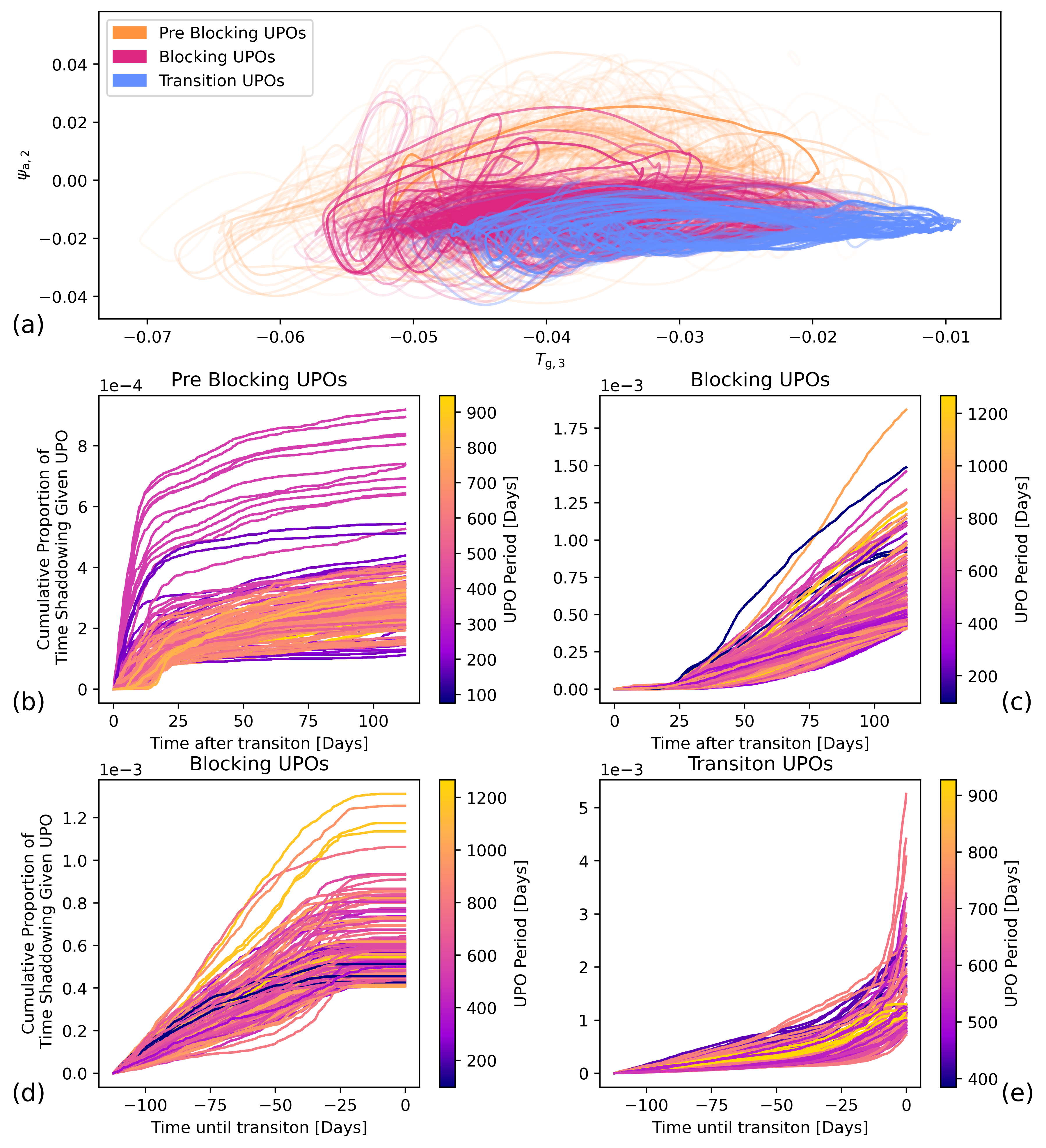}
        \caption{Figure (a) shows the three sets of UPOs, colour coded by which set they are in, the shade of the colour displays the magnitude of cumulative shadowing (CS) overall, with darker shades representing UPOs that are shadowed more frequently. (b) each curve represents a UPO, and the y-axis shows the proportion of time that the UPO shadows the trajectory. The shading is the UPO period. This subplot shows the CS of the UPOs in the Pre-Blocking set (those in orange in (a)), for the first 100 days of the time in the cluster. (c) same as (b), but showing the CS of the Blocking set of UPOs, shown in pink in (a). (d) shows the CS of the set of Blocking UPOs for the final 100 days before transitioning. (e) shows the same as (d), but for the Transition set of UPOs, shown in blue in (a).}
        \label{fig:LHS_upo_prediction_collection}
    \end{figure}

    Looking at trajectories that spend long time periods (at least 110 days) in only one cluster, we found at least three distinct sets of UPOs: one where the the cumulative shadowing (CS) increases primarily when the trajectory first enters the cluster (called Pre-Blocking), another set where the CS mainly increases before a transition (called Transition), and a final set where the CS mainly increases in between entering and transitioning from the regime (called Blocking). We show these three distinct sets of UPOs for the LHS cluster in Figure~\ref{fig:LHS_upo_prediction_collection} (corresponding figures for the RHS can be found in Section~\ref{app:rhs_figs}). We set a minimum time for the trajectory to spend in one cluster because, as we showed in Section~\ref{sec:results_sub:transitions}, trajectories that do not spend enough time in either cluster tend not transition through specified regions, meaning that one cannot predict these transitions with only three groups of UPOs. 

    \begin{figure}
        \centering
        \includegraphics[width=\linewidth]{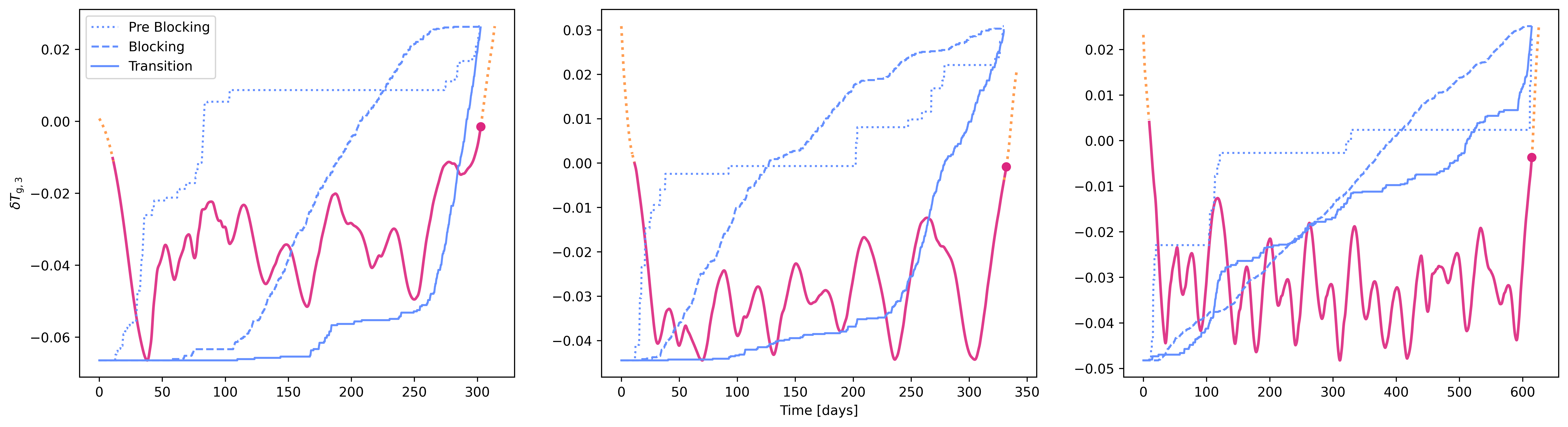}
        \caption{Examples of cumulative shadowing (CS) of three trajectories in the left hand cluster. A time series of $\QgsGroundTemp[3]$ is shown in pink when in the left hand cluster and orange when in the other cluster, a dot is shown where this transitions to the RHS cluster. The solid blue line shows the cumulative shadowing of the transition set, the dashed line shows the CS of the set associated with blocking, and the dotted line shows the CS of the pre-blocking set. We see before the transition the blocking set flattens out, while the CS of the transition set begins to increase considerably. All three CS curves are normalised and scaled for visualisation purposes.}
        \label{fig:transition_methodology}
    \end{figure}

    In Figure~\ref{fig:transition_methodology} we show three time series of the trajectory, and the CS for each one. We see in these examples that before the transition there is a clear increase in the CS of the Transition set of UPOs, while the CS of the Blocking set plateaus. In addition, we show the CS of the Pre-Blocking set shows that largest increase before the trajectory settles onto the set of Blocking UPOs. This appears to show that this cumulative shaddowing method could provide a useful early warning indicator for transitions between clusters of the attractor.
    
\section{Conclusions}\label{sec:conclusions}
    In this paper, identification of unstable periodic orbits and continuation methods are used to explain the dynamical behaviour of a reduced order land-atmosphere model. While the use of UPOs to characterise atmospheric blocking has been studied before~\citep{gritsun2008,lucarini2019a}, in both barotropic and baroclinic models, this study has built on this by using the UPOs to cluster the attractor into specific regions. These clusters were then used to identify transition regions, and link these with atmospheric behaviour.
    
    The clusters in this model are of interest as they represent areas where the ground heat shows an anomaly in either the east or west regions of the domain. Through the heat transfers between the land and the atmosphere, this anomaly causes, on average, higher temperatures in the atmosphere, which can reinforce atmospheric blocking situations, that result from Rossby wave deflections induced by the orography.

    This paper has also introduced a significant modification of the \textsf{qgs} framework~\citep{demaeyer2020}, which uses symbolic Python to construct the model equations. This modification allowed us to export the model equations in Julia (the language used for numerically finding the UPOs) and Fortran (formatted for using in the AUTO continuation software). In addition, it facilitated exporting certain parameter values as variables.
    
    Through the use of continuation software we have also been able to provide a more comprehensive understanding of the origin of the structure of the attractor that lead to atmospheric blocking. We found the paths to chaos from two stable orbits (for parameter values found in~\citet{xavier2024}).
    
    Finally, we explored the possibilities offered by the concept of shadowing of UPOs~\citep{maiocchi2022} to define cumulative shaddowing (CS), and identification of three sets of UPOs associated with different times of the blocking regime lifecycle. As the accuracy of numerical weather prediction's ability to forecast the onset and decay of atmospheric blocks still lags behind other features of the atmosphere, other methods that can forecast the onset or decay are of particular interest. While the approach that we used here is somewhat simplistic, we believe that proximity to particular UPOs or the rate of change of the CS could be useful as a training parameter in a machine learning, or more traditional forecasting approach, for forecasting transitions in state space.

    Shadowing has been used with success in simple models to understand transitions~\citep{maiocchi2024}, and we believe that there is untapped potential to do similar for more complex models. In higher dimensional models, however, it could be more fruitful to analyse the shadowing of collections of UPOs rather than individual UPOs. This is because in higher dimensional systems we will usually have fewer UPOs to study, and there are a larger number of unstable directions that the trajectory could diverge from the UPO.

    Future work is required to investigate if similar techniques to those shown in this paper can expand to more complex models that more accurately model atmospheric blocking. In addition with the increase of global temperatures, and the change in north-south temperature gradients as a result of anthropogenic climate change, such models and techniques could help understand how atmospheric blocking events will differ in the future, a topic that models and observations still differ largely on~\citep{riboldi2020, davini2020}.

    \begin{acknowledgments}
    This project has received funding from the European Union’s Horizon 2020 research and innovation programme under the Marie Sklodowska–Curie grant agreement no. 956170. In addition, funding has been provided through the ``Fédération Wallonie-Bruxelles" with the instrument ``Fonds Spéciaux de Recherche". The authors would like to thank Anupama K Xavier for providing results from previous studies and for helpful discussions.
\end{acknowledgments}

\section*{Data Availability Statement}
    The code used to numerically calculate the unstable periodic orbits for this paper will be shared on GitHub once documentation is complete (before the end of the review process). Reviewers can contact the authors directly for a copy of the code. Bifurcation analysis was undertaken using references~\citet{demaeyer2025, demaeyer2025_zenodo}. The qgs framework is available from reference~\citet{demaeyer2020}, the modified version of qgs (v1.0) is being prepared to be included in the GitHub repository before the end of the review process.

    \appendix
    \section{Numerical Methods}\label{app:numerical_methods}
\subsection{Newton Raphson Method}
    Given a system of ordinary differential equations $\dot{\IC[]}=f(\IC[])$. We want to solve the equation: $\systemFlow[\UPOperiod](\IC)-\IC=0$.
    This is a system of $N$ equations, where we have $N+1$ unknowns; $N$-dimensional space unknowns in space, and the period $\UPOperiod$. Given an initial guess $(\IC, \UPOperiod_0)$, which we discuss how to find in Section~\ref{sec:methods_sub:ic}, we define the $i$-th iteration of the variables $(\IC[i], \UPOperiod_i)$. The algorithm aims to calculate corrections $(\Delta \IC[i], \Delta \UPOperiod_i)$ to improve the initial guesses, in other words:

    \begin{equation*}
        \left\|\systemFlow[\UPOperiod_i+\Delta \UPOperiod_i]\left(\IC[i]+\Delta \IC[i]\right)-\left(\IC[i]+\Delta \IC[i]\right)\right\|<\left\|\systemFlow[\UPOperiod_i]\left(\IC[i]\right)-\IC[i]\right\|,
    \end{equation*}

    These correction terms are calculated by $x_{i+1}=\IC[i]+\Delta \IC[i]$ and $\UPOperiod_{i+1}=\UPOperiod_i+\Delta \UPOperiod_i$. Expanding the original problem using a Taylor series:
    \begin{equation}\label{eq:newton_linearisaton}
        \begin{split}
            \system[\UPOperiod_i+\Delta \UPOperiod_i]&(\IC[i]+\Delta \IC[i])-
            (\IC[i]+\Delta \IC[i]) \\
            &\approx \systemFlow[\UPOperiod_i](\IC[i])+\Delta \IC[i]\partialDeriv{\systemFlow[\UPOperiod_i](X)}{X}\bigg|_{X=\IC[i]}+\Delta \UPOperiod_i\partialDeriv{\systemFlow[\UPOperiod](\IC[i])}{\UPOperiod}\bigg|_{\UPOperiod=\UPOperiod_i}-\bigg(\IC[i]-\Delta \IC[i]\partialDeriv{X}{X}\bigg)\\
            &\approx \systemFlow[\UPOperiod_i](\IC[i]) - \IC[i]+\Delta \IC[i]\bigg(\partialDeriv{\systemFlow[\UPOperiod_i](X)}{X}\bigg|_{X=\IC[i]} - \idMat\bigg)+\Delta \UPOperiod_i\partialDeriv{\systemFlow[\UPOperiod](\IC[i])}{\UPOperiod}\bigg|_{\UPOperiod=\UPOperiod_i}
        \end{split}
    \end{equation}

    Where $\idMat$ is the identity matrix and $\partialDeriv{\systemFlow[\UPOperiod_i](X)}{X}$ is the tangent linear operator. The term $\partialDeriv{\systemFlow[\UPOperiod_i]}{\UPOperiod}(\IC[i])$ is the derivative of the solution of $\dot{\IC[]}=f(\IC[])$ w.r.t time, which is evaluated at the iteration. This is equivalent to $f(\systemFlow[\UPOperiod_i](\IC[i]))$.

    To reduce the number of degrees of freedom, the phase condition is added that requires the orbit and the correction vector to be orthogonal to one another:
    \begin{equation*}
        f(\systemFlow[\UPOperiod_i](\IC[i]))\cdot \Delta \IC[i]=0
    \end{equation*}

    Together, this becomes a problem involving $N+1$ unknowns with $N+1$ equations and can be expressed as:
    \begin{equation}\label{eq:newton_lin_system}
        \left[\begin{array}{cc}
            \partialDeriv{\systemFlow[\UPOperiod_i](X)}{X}-\idMat & f\left(\systemFlow[\UPOperiod_i]\left(\IC[i]\right)\right) \\
            \transpose{\left(f\left(\IC[i]\right)\right)} & 0
        \end{array}\right]
        \left[\begin{array}{c}
            \Delta \IC[i] \\
            \Delta \UPOperiod_i
        \end{array}\right] = 
        \left[\begin{array}{c}
            \IC[i] - \systemFlow[\UPOperiod_i]\left(\IC[i]\right) \\
            0
        \end{array}\right]
    \end{equation}

    This iteration is continued until the following two conditions are smaller than a specified magnitude:
    \begin{itemize}
        \item $|\systemFlow[\UPOperiod_i](\IC[i])-\IC[i]|<\Delta_a$
        \item $|(\Delta \IC[i], \Delta \UPOperiod_i)|<\Delta$
    \end{itemize}

    This algorithm needs slight editing for when the differential system includes integral terms where another condition needs to then be added, which causes the linear system to no longer include a square matrix. To solve this~\citet{abad2011} suggests using the least norm method and singular value decomposition to solve the non-square system. This is not a concern in our case.

    Once a correction step is calculated, a line search optimisation step is used to choose the largest step that should be taken in the direction $(\Delta \IC[i], \Delta \UPOperiod_i)$~\citep{gritsun2008}.

\subsection{Tensor Correction}
    We first construct and then solve the linear system, as discussed in the Newton Raphson method outlined above:
    \begin{equation}\label{eq:newton_system}
        A \cdot \left[\begin{array}{c}
            \Delta \IC[i] \\
            \Delta \UPOperiod_i
        \end{array}\right] = \left[\begin{array}{c}
            \IC[i] - \systemFlow[\UPOperiod_i]\left(\IC[i]\right) \\
            0
        \end{array}\right]
    \end{equation}
    where $A$ is the matrix shown in Equation~\ref{eq:newton_lin_system}. Issues arise when the matrix $A$ is close to a singular matrix. Such cases are a symptom of the linearisation assumption breaking down for certain directions in phase space. To overcome this, the resulting step size of the algorithm in these directions may need to be very small. Thus leading to an increase in the time taken to reach an optimum solution.
    
    To overcome these issues~\citet{gritsun2008} proposed using a tensor correction method~\citep{schnabel1984, bader2005}, where the second order non-linear terms are included in the Taylor series. To do this we expand the orininal equaion as in Equation~\ref{eq:newton_linearisaton}. To simplify how this is written we will first introduce some new notion, note we have slightly modified the notation found in~\citet{gritsun2008}. We construct a $N+1$ dimensional vector $\CombinedVector=[\IC[i], \UPOperiod_i]$ and a $N+1$ dimensional vector $\Delta \CombinedVector = [\Delta \IC[i], \Delta \UPOperiod_i]$, the latter designating the correction step. 
    
    We also define a function $F:\mathbb{R}^{N+1}\mapsto \mathbb{R}$ as the difference between the initial condition, and the integrated initial condition. In other words $F$ measures the gap between the starting and finishing position of the estimated UPO: $F(\IC[i], \UPOperiod_i)=F(\CombinedVector) = \systemFlow[\UPOperiod_i]\left(\IC[i]\right)-\IC[i]$. Also, as we are now primarily working with the vector $\CombinedVector$, the dimension of the following system will be $N+1$, where $N$ is the dimension of the phase space. Finally taking the Taylor series to the second order we obtain:

    \begin{equation}
        \begin{split}
            F(\CombinedVector[i+1])&=F(\CombinedVector+\Delta \CombinedVector)\\
            &\approx  F(\CombinedVector)+\partialDeriv{F(\CombinedVector)}{[\CombinedVector]_j}[\Delta \CombinedVector]_j+\frac{1}{2}\partialDeriv{}{[\CombinedVector]_j}\partialDeriv{F(\CombinedVector)}{ [\CombinedVector]_k}[\Delta \CombinedVector]_k[\Delta \CombinedVector]_j
    \end{split}
    \end{equation}
    Where $[\CombinedVector]_j$ denotes the $j$-th element of the vector of the $i$-th iteration of the algorithm.

    As in Equation~\ref{eq:newton_system} we can express the equation in matrix form:
    \begin{equation}\label{eq:tensor_correction}
        A\Delta \CombinedVector+\frac{1}{2}\partialDeriv{}{[\CombinedVector]_j} [A\Delta \CombinedVector][\Delta \CombinedVector]_j=-F(\CombinedVector)
    \end{equation}

    Again, the reason we are including the tensor correction is that the matrix $A$ could be close to singular. So we focus on the columns of the singular value decomposition $A=U\Sigma \transpose{V}$ that are close to zero, and which are causing the issue. The decomposition sorts the diagonal, or `singular', values of $\Sigma$ so that they are decreasing in magnitude. Now we are interested in the singular values which are sufficiently small, which will occur in the last columns of the matrix. Let's assume that the first $K$ singular values are `large enough', meaning that the last $N+1-K$ values are too close to zero. We start by multiplying Equation~\ref{eq:tensor_correction} by $\transpose{U}$, and making the substitutions $\transpose{V} \Delta \CombinedVector = l$ and $-\transpose{U} F(\CombinedVector)=g$:

    \begin{equation}\label{eq:tensor_correction_simplification}
        \Sigma l+\transpose{U}\frac{1}{2}\partialDeriv{}{ [\CombinedVector]_j} (AVl)[Vl]_j=g
    \end{equation}

    We can separate the solution into an orthogonal sum of vectors $l=l_1+l_2$ which correspond to the basis of the matrix $V$. \citet{gritsun2008} found that for numerically finding UPOs, it was sufficient to take $K=N-1$, in other words only, focus on the final two singular values of the system. This reduces the problem into a linear system of dimension $N-1$, and a non-linear problem of dimension $2$. This means that the two orthogonal vectors will look like: $l_1=(v_1, \dots, v_{N-1}, 0, 0)$ and $l_2 = (0, \dots, 0, v_{N}, v_{N+1})$. Therefore we can solve the first $N-1$ linear equations to obtain $l_1$ by taking $[l_1]_k=[\Sigma^{-1}]_k~g_k$ for $k=1,\dots,N-1$. We then substitute $l_1$ into the following two non-linear equations:

    \begin{equation}
        \begin{split}
            \sigma_N [l_2]_N + \bigg(u_N, \frac{1}{2}\partialDeriv{}{[\CombinedVector]_j}&(A(Vl_1+v_n[l_2]_N+v_{N+1}[l_2]_{N+1}))\\
            &[Vl_1+v_n[l_2]_N+v_{N+1}[l_2]_{N+1}]_j\bigg)=[g]_N\\
            \sigma_{N+1} [l_2]_N + \bigg(u_{N+1}, \frac{1}{2}\partialDeriv{}{[\CombinedVector]_j}&(A(Vl_1+v_n[l_2]_N+v_{N+1}[l_2]_{N+1}))\\
            &[Vl_1+v_n[l_2]_N+v_{N+1}[l_2]_{N+1}]_j\bigg)=[g]_{N+1}
        \end{split}
    \end{equation}
    where $\sigma_i,~ v_i,~u_i$ are the $i$-th column of the matrices $\Sigma,~V$, and $U$ respectively. The non-linear terms in both equations are the same, but they are projected onto the direction of $u_N$ or $u_{N+1}$.

    To calculate the non-linear terms we have to calculate the directional derivatives:
    \begin{equation*}
        \partialDeriv{}{[\CombinedVector]_j}(A\alpha)[\beta]_j = \partialDeriv{}{\beta}(A\alpha)
    \end{equation*}
    Where $\alpha, \beta\in[Ql_1,~v_n,~v_{N+1}]$ there are therefore nine permutations possible. For example if we take $\alpha=v_N,~ \beta=v_{N+1}$ we can approximate the derivative by taking a finite difference approximation:
    \begin{equation*}
        \partialDeriv{}{v_{N+1}}(Av_N) \approx\frac{1}{\delta}\bigg(A(u_i+\delta v_{N})v_{N+1} - A(u_i)v_{N+1}\bigg)
    \end{equation*}
    The calculation of $A(\CombinedVector+\delta v_{N})v_{N+1}$ is equivalent to integrating the tangent linear model, as in Equation~\ref{eq:newton_linearisaton}. It consists in integrating the tangent linear model from an initial condition $[\IC[i]]_j+\delta [v_{N}]_j$ (for $j\in[1, N]$) for a time of $\UPOperiod_i+\delta [v_{N}]_{N+1}$.
    
    We already have the value of $A(\CombinedVector)$ from the construction of the linear system. As we have nine different permutations for $\alpha$ and $\beta$ this boils down to doing nine additional integrations of the tangent linear model.
    
    We then solve the two non-linear simultaneous equations numerically. There could be up to four possible solutions for $([l_2]_{N}, [l_2]_{N+1})$ as the solutions are the intersection of two parabolas, and we pick between these solutions by picking the one that results in the lowest numerical error of the non-linear system. This gives us the full vector $l=l_1+l_2$, and in turn the next correction step: $\Delta \CombinedVector=Vl$. The line search step is used to ensure we take the magnitude of the correction $(\Delta \IC[i], \Delta \UPOperiod_i)$ which minimises $F(\CombinedVector[i+1])$.

    This method clearly increases the run time for each step (by approximately an order of 10), so we limit the use of the tensor correction to only steps where the matrix $A$ has close to $0$ singular values.

    \section{UPO shadowing - RHS figures}\label{app:rhs_figs}
    Here we display the three set of UPOs for the RHS cluster, as shown for the LHS in the main paper in Section~\ref{sec:results_sub:upo_shaddowing}. 

    \begin{figure}
        \centering
        \includegraphics[width=0.8\linewidth]{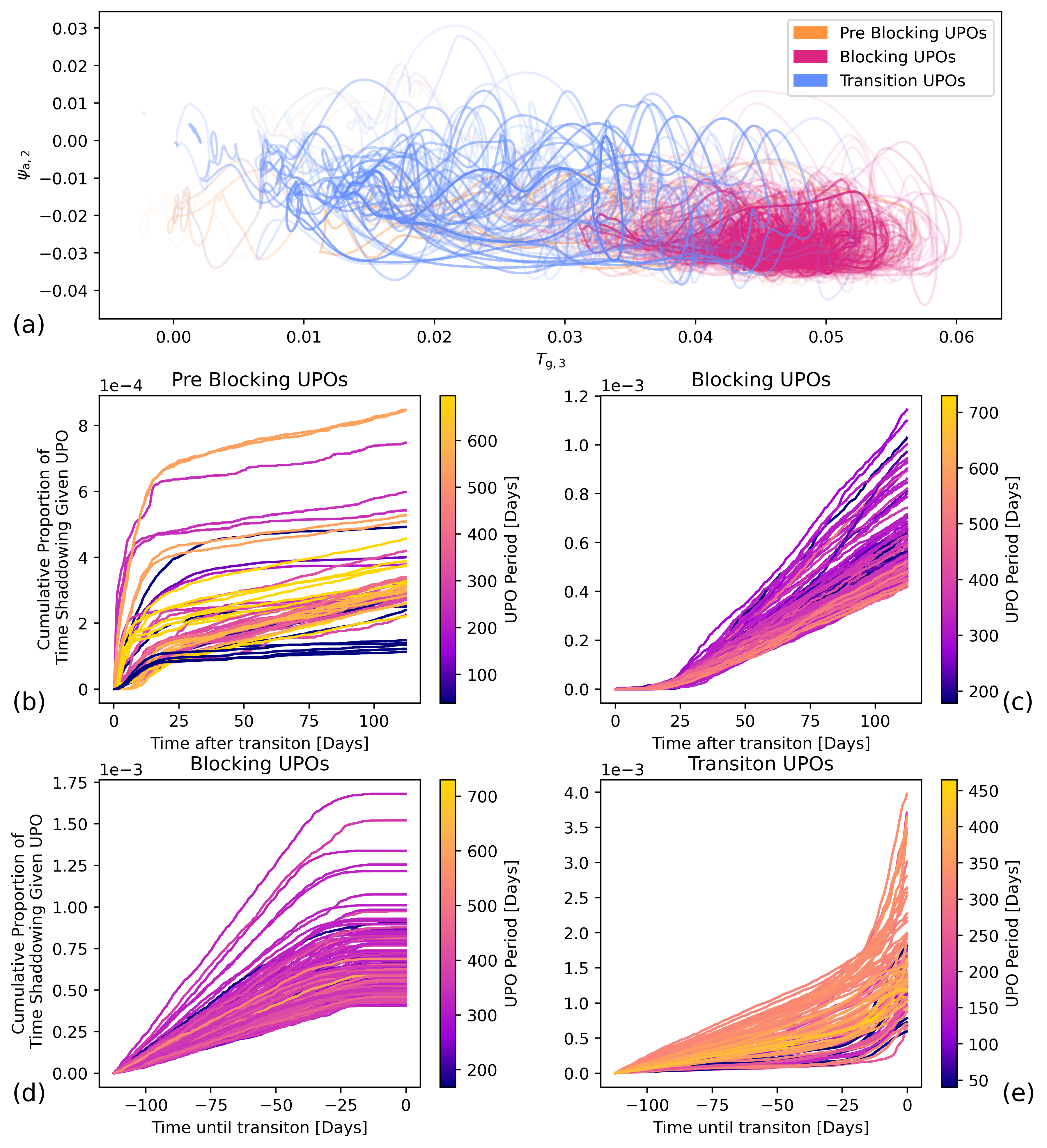}
        \caption{Same as shown in Figure~\ref{fig:LHS_upo_prediction_collection} but for the RHS cluster. Figure (a) shows the three sets of UPOs, colour coded by which set they are in, the shade of the colour displays the magnitude of cumulative shadowing (CS) overall, with darker shades representing UPOs that are shadowed more frequently. (b) each line represents a UPO, and the y-axis shows the proportion of time that the UPO shadows the trajectory. The shading is the UPO period. This subplot shows the CS of the UPOs in the Pre-Blocking set (those in orange in (a)), for the first 100 days of the time in the cluster. (c) same as (b), but showing the CS of the Blocking set of UPOs, shown in pink in (a). (d) shows the CS of the set of Blocking UPOs for the final 100 days before transitioning. (e) shows the same as (d), but for the Transition set of UPOs, shown in blue in (a).}
        \label{fig:RHS_upo_prediction_collection}
    \end{figure}

    \bibliography{MyBib}
\end{document}